\documentclass[10pt,preprint]{aastex}	

\usepackage{natbib}

\DeclareGraphicsExtensions{.pdf,.png,.jpg}

\newcommand{\mum}{$\mu$m}
\newcommand{\etal}{et~al.}
\newcommand{\vsini}{$v \sin i$ }

\begin{document}
\bibliographystyle{plainnat}

\title{Exploring the Effects of Stellar Rotation and Wind
Clearing: \\ Debris Disks Around F Stars}

\slugcomment{Accepted to AJ, \today}
\shortauthors{Mizusawa et al.}
\shorttitle{F Star Debris Disks}

\author{Trisha F.\ Mizusawa\altaffilmark{1,2},
Luisa M.\ Rebull\altaffilmark{3}, 
John R.\ Stauffer\altaffilmark{3},\\
Geoffrey Bryden\altaffilmark{1,4},
Michael Meyer\altaffilmark{5},
Inseok Song\altaffilmark{6}
}

\altaffiltext{1}{Work conducted while resident at: NASA Star and
Exoplanet Database (NStED), 1200 E.\ California Blvd., California
Institute of Technology, Pasadena, CA 91125}
\altaffiltext{2}{Currently Florida Institute of Technology, 150 W.
University Boulevard, Melbourne, FL 32901; trisha.mizusawa@gmail.com}
\altaffiltext{3}{Spitzer Science Center (SSC), 1200 E.\ California
Blvd., California Institute of Technology, Pasadena, CA 91125}
\altaffiltext{4}{Jet Propulsion Laboratory, 4800 Oak Grove Blvd,
Pasadena, CA 91109}
\altaffiltext{5}{ETH, Zurich, Switzerland}
\altaffiltext{6}{Department of Physics and Astronomy, University of
Georgia, Athens, GA 30602}

\begin{abstract}

We have conducted a study of debris disks around F stars in order to
explore correlations between rotation, stellar winds, and
circumstellar disks. We obtained new  24 \mum\ photometry from
Spitzer's  Multiband Imaging Photometer for Spitzer (MIPS) camera for
a sample of 188 relatively nearby  F dwarfs with various rotation
rates and optical colors, and combined it with archival MIPS data for
66 more F stars, as well as Wide-field Infrared Survey Explorer (WISE)
data for the entire sample, plus 9 more F stars.   Based on the
objects' $K_s-[24]$ and $[3.4]-[22]$ colors, we identify 22 stars in
our sample  as having 22 and/or 24 \mum\ excesses above our detection
limit, 13 of which are new discoveries. Our overall disk detection
rate is 22/263, or 8\%, consistent with previous determinations of
disk fractions in the Solar neighborhood.   While fast rotating stars
are expected to have strong winds capable of efficiently removing
dust, we find no correlation between rotational velocity and infrared
excess. Similarly, we find no significant difference in excess
detection rate between late-type F stars, which have convective
surfaces, and early-type F stars, which have fully radiative
envelopes. However, the essentially unknown range of ages in this
sample may be washing out any effects relating rotation, winds, and
disks.

\end{abstract}


\section{Introduction}
\label{sec:intro}

Infrared Astronomy Satellite (IRAS) and Spitzer Space Telescope
(Werner \etal\ 2004) observations of field A stars and of open
clusters have shown that debris disks are common at early ages and
decrease in both frequency and luminosity for older stars.   Based on
images from the Multiband Imaging Photometer for Spitzer  (MIPS; Rieke
\etal\ 2004), Su et al.\ (2006) found about a 30\% frequency of such
disks at 24 $\mu$m for a sample of $<$ 1 Gyr old A stars, with the
upper envelope to the 24 $\mu$m excess flux decaying roughly as
$t_0/t$, with $t_0 \sim$ 150 Myr.  Studies of a number of open
clusters (e.g., Currie et al.\ 2009 and references therein) agree with
the general trends found for the field A dwarfs.  

The debris disk frequency for low mass stars is less well-determined,
but again observations of field stars (e.g., Carpenter et al.\ 2009)
and of open clusters (e.g., Rebull et al.\ 2008 and references
therein) indicate that the debris disk frequency and luminosity peaks at
young ages and declines after that, with few or no debris disks
detected at 24 $\mu$m by Hyades or Praesepe age ($\sim$ 650 Myr; Cieza
et al.\ 2008, Urban \etal\ 2012).  

For the relatively luminous A stars, radiation pressure removes dust
particles from the disk on timescales $<<$ 1 Myr; the dust in these
debris disks must be constantly resupplied from collisional events
amongst planetesimals. The lower luminosities of late-type (GKM)
dwarfs are less effective in removing the dust via blowout or 
traditional (radiative) Poynting-Robertson drag. However, as described
by Chen et al.\ (2005) and Plavchan et al.\ (2005, 2009), winds from
rapidly rotating, low mass stars should engender a particulate
Poynting-Robertson effect that is more efficient at removing orbiting
dust. 

F stars inhabit a sweet-spot between the realms of more massive and
less massive stars. It is within the F spectral type where outer
convective envelopes begin to appear, at spectral type F5. Higher mass
stars have convective cores and radiative envelopes, and winds from
the photonic Poynting-Robertson effect scour dust from circumstellar
disks.  Lower mass stars have strong dynamo activity caused by
convective motions and magnetic fields in their outer convective
envelopes; strong winds should scour dust from circumstellar disks. 
As shown by Simon \& Landsman (1991), Wolff \& Simon (1997), and
others, there is good evidence that the transition to strong, dynamo
driven winds occurs for stars later than F5 -- with activity
indicators such as \ion{Ca}{2} HK emission correlated with rotation
for later type stars, and not correlated for earlier type stars. 

It has long been known that the \vsini distribution of main sequence
stars shows a transition at spectral type F.  Early type (O, B, and A)
stars are generally rapid rotators, with average rotational velocities
of 200 to 300 km s$^{-1}$.   Main sequence G, K, and M dwarfs (older
than a few hundred Myr) are generally very slow rotators, with mean
rotational velocities of only a few km s$^{-1}$.  There is thus a
sharp drop in the mean rotation rate of main sequence (MS) stars as one
goes from F0 to F9.  Interestingly, when this facet of rotation on the
MS was first recognized, two possible explanations were advanced --
angular momentum loss via winds (Mestel \& Roxburgh 1962) and angular
momentum redistribution into planetary systems (Huang 1965).  The
former explanation was eventually recognized as the primary physical
mechanism responsible for the drop in rotational velocities. 

However, the strength of those winds -- or at least the amount of
angular momentum they carry away -- is a subject of controversy. Some
models (e.g., Denissenkov \etal\ 2010, Sills \etal\ 2000) predict that
only the outer convective envelope is spun down (initially) by the
wind; because the amount of mass (and the moment of inertia) present
in the outer convective envelope of late F stars is very small, in
this event the inferred mass loss rate could be small even for rapidly
rotating F stars.  Other models, however, predict strong coupling
between the convective envelope and the radiative interior, and
postulate that the entire star is spun down on short timescales even
for late F stars (Bouvier \etal\ 1997).  The inferred mass loss rate
must be much higher in this event.  

Because of the rapid change of wind properties through the F star
regime, stars of these masses provide unique laboratories for studies
of wind-related effects.   Debris disks could be more common around F
stars near the transition between early and late F stars, and could
persist for longer periods than for either higher or lower mass stars
(and may result in planets with high obliquities; see Winn et al.\
2010).  However, for stars with massive disks, the effect of PR drag
may be small compared to mutual collisions of dust within a disk such
that the grains are ground down into fine enough particles that
radiation pressure can remove the dust (e.g., Wyatt 2005).

We have conducted a MIPS 24 $\mu$m survey of 254 field F dwarfs, half
of which rotate quickly ($>$15 km s$^{-1}$), the other half of which
rotate slowly ($\leq$15 km s$^{-1}$). We added to the sample 9 objects
detected in WISE but missing from the MIPS archive. Our goal was to
determine if (a) F dwarfs have a comparatively high debris disk
frequency and (b) we see evidence (via debris disk frequency) for
development of winds at the F5 boundary where solar-type winds are
expected to develop.  Section~\ref{sec:analysis} presents the data
analysis and reduction, \S\ref{sec:disks} identifies infrared
excesses, and \S\ref{sec:results} discusses the results from this
paper.

\section{MIPS 24 \mum\ Photometry and Data Analysis}
\label{sec:analysis}

\subsection{Target selection}
\label{sec:target}

We merged the Hipparcos catalog with other catalogs containing
spectroscopic rotational velocities (see below). We selected all stars
with spectral types F0 to F9 and distances less than 100 parsecs, and
required that the targets were above (or below) a galactic latitude of
9$\arcdeg$ ($\mid b\mid> 9\arcdeg$) and had an ecliptic latitude above
19$\arcdeg$ ($\mid\beta\mid>19\arcdeg$).  This resulted in $\sim$7000
stars.  In order to reinforce the spectral type restriction, we
retained stars with $0.30  > B-V > 0.52$. To retain only main sequence
stars and mitigate somewhat against binaries, we eliminated stars more
than 0.1 mag fainter than the single-star ZAMS (Zero-Age Main
Sequence) or brighter than 0.6 mag above the ZAMS. Because we were
interested in correlations with rotation, we only retained objects for
which there were measurements of $v \sin i$.  We split the sample into
a blue set ($0.30 \leq B-V \leq 0.41$) and a red set ($0.43 \leq B-V
\leq 0.52$). Note that there is a clear and artificially-imposed
distinction in color for these two samples; we dropped stars with
$B-V$ colors between 0.41 and 0.43 (e.g., colors centered on F5) to
make the samples more cleanly and explicitly divided into early and
late F (radiative and convective envelopes, respectively).  Each of
these sets was again bifurcated into a slowly rotating set and a
rapidly rotating set (\vsini $\leq$ or $>$ 15 km s$^{-1}$).  This
division between fast and slow rotation was selected because the X-ray
flux saturates for G stars at about 15 km s$^{-1}$; we assume that
wind strength and X-ray flux are roughly correlated. Lacking other
guidance, we extrapolated that limit to the F star regime. Each group
of remaining F stars was then sorted by distance, which we then
compared with the objects already observed by Spitzer. We requested
new Spitzer observations of the $\sim50$ closest objects not yet
observed in each group. 

Our target selection for new observations by definition deliberately
sought objects that were not already in the Spitzer archive. There is
some reason to believe that observations already obtained by others in
the Spitzer archive may be biased {\em towards} objects with disks --
for example, Mo\'or \etal\ (2011) which deliberately observed F stars
thought to have disks, or Rebull \etal\ (2008) which deliberately
observed stars thought to be members of the Beta Pic Moving Group
(which includes an F star, and stars belonging to this moving group
are young enough that they are still likely to have substantial
disks). We must include in our analysis objects whose observations
were already in the Spitzer archive, but we do so knowing that those
archive observations may not have been taken with the same sensitivity
as our new observations. All of our targets are bright (easily
detected), so this is not likely to introduce any biases, per se. 
However, there were considerably more red (late) F stars already in
the Spitzer archive than blue (early) F stars (49/156, vs.\ 18/109),
so when the entire sample is considered, the net number of late F
stars in the Spitzer archive (our new $\sim$100 observations of red F
stars plus that of others, total of 156 objects) is greater than the
number of early F stars (total of 109 objects). It is also true that
the late F stars are on average closer to us than the early F stars.
As a result, our blue sample reaches out to greater distances (96 pc)
than the red sample (67 pc). 

Subsequent to our original proposal, the all-sky catalog from the
Wide-field Infrared Survey Explorer (WISE; Wright \etal\ 2010) has
been released. WISE surveyed the whole sky in four bands, 3.4, 4.6,
12, and 22 microns.  WISE is very different than MIPS -- the WISE 22
\mum\ bandpass is distinctly different than the MIPS 24 \mum\
bandpass, the spatial resolution is twice as good with MIPS
($\sim$6$\arcsec$) as with WISE ($\sim$12$\arcsec$), and the
sensitivity is in general better with MIPS than with WISE (where the
sensitivity varies with ecliptic latitude). However, since WISE
observed the whole sky, and since our targets are generally bright, we
can use the WISE data to complement our sample.  For example, we can
use the [3.4]$-$[22] (W1-W4) color as an independent check on the IR
excess we determine using $K_s-$[24]. In some cases, we can include
objects that were dropped from our Spitzer target selection because we
thought they were in the Spitzer Archive, but the objects were just
off the edge of the observations.   

We selected targets for our sample (for our new observations or from
the Spitzer archive, or from WISE) without particular bias towards
presence or absence of disks. Because we started from many different
catalogs aimed at studies of nearby stars, there should be no
particular biases imposed by, e.g., requiring $v \sin i$ any more than
the requirement that there be a good $B$ and $V$ measurement. 
However, as will be seen below, we probably have an implicit bias in
that the rapid rotators are more likely to be young. 

\begin{figure}[h]
    \plotone{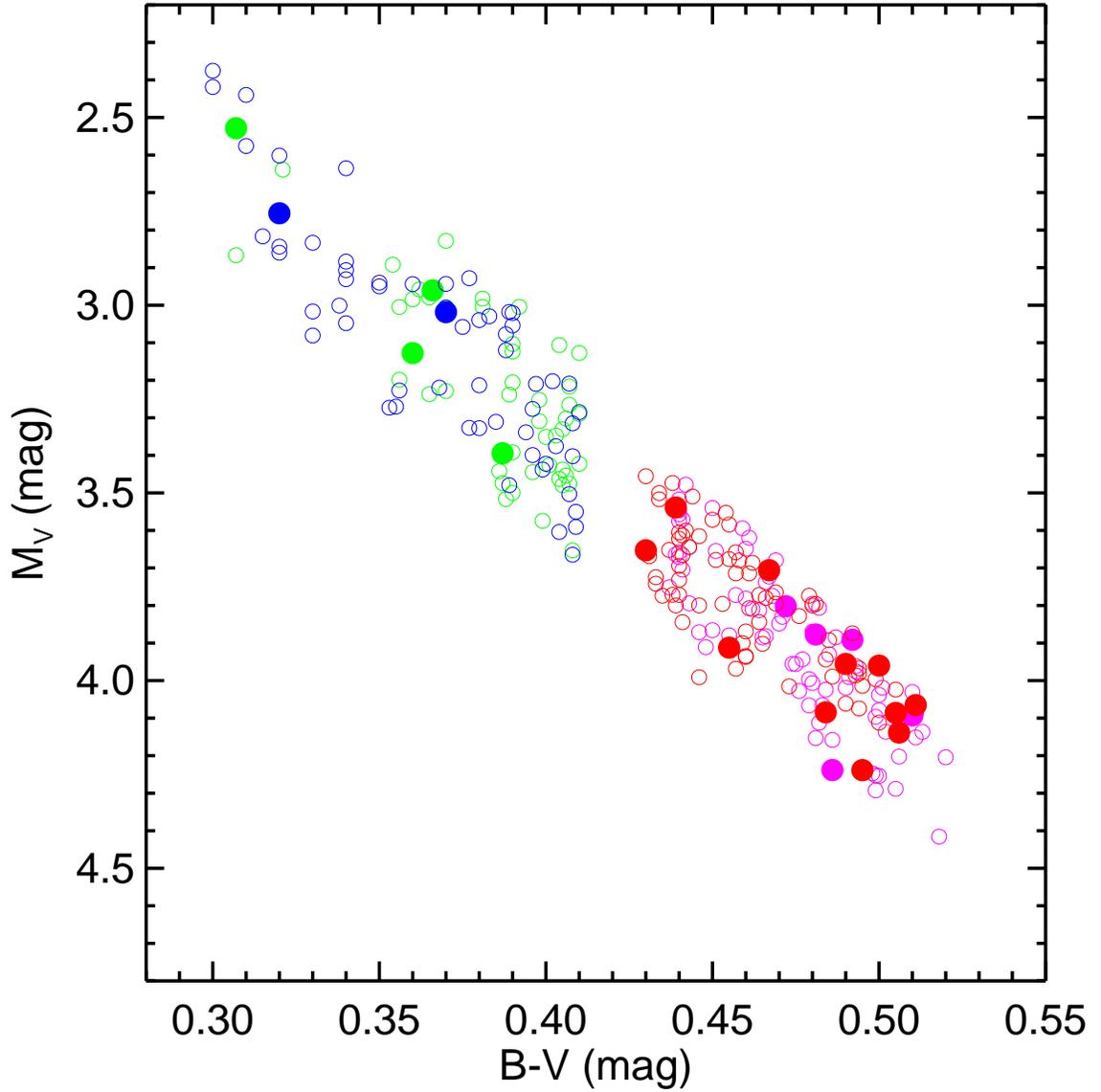}
\caption{$M_v$ vs. $B-V$ for the full sample of stars, which shows the
as-designed gap in colors near $B-V\sim$0.42, or type F5.  Green
circles represent the blue slow group, blue circles represent the blue
rapid groups, magenta circles represent the red slow group, and red
circles represent the red rapid group.  Filled symbols indicate
infrared excess stars (see section \ref{sec:disks}). In this Figure,
there is no statistical difference between the distributions of
either of the red groups or either of the blue groups, suggesting that
the stars span the same range of ages.  However, the age range may
be significant; see discussion in \S\ref{sec:results}.}
  \label{MvBV}
\end{figure}

The final set of stars is shown in a $M_v$ vs.\ $B-V$ color-magnitude
diagram (CMD) in Figure~\ref{MvBV} (and listed in Table~\ref{fulltab};
also see Table~\ref{tab:disks} for relevant notes about certain
stars).   The visible gap between the blue and red  samples in
Figure~\ref{MvBV} is an explicit result of the selection process
above; this gap corresponds to the F5 division between stars with
radiative and convective envelopes.   We could not identify any
statistically significant difference between the Fig.~\ref{MvBV}
distribution of the blue-rapid and blue-slow stars, nor in the
distribution of red-rapid and red-slow stars, which would imply
similar age distributions.  However, there is an expectation that the
rapid rotators should be younger than the slow rotators, on average,
and this effect should be more pronounced for the red sample; we will
discuss this in more detail below.  
Any ages assumed from the $M_v$/$B-V$ CMD are unlikely to be very
accurate.

\subsection{Observations}
\label{sec:obs}

Each of the 188 targets observed as part of our original Cycle-5
program  50449 (P.I.: L.\ Rebull) was observed using a MIPS photometry
AOR (Astronomical  Observation Request) at 24 \mum.  Our AORs are the
same for all of our targets, using  one cycle of 3 seconds integration
time at 24 \mum\ for a total integration  time of 42 sec pixel$^{-1}$
on-target.  There are 69 more objects for which MIPS-24 observations
were obtained external to program 50449.   While most of these were
MIPS photometry AORs targeting our object of interest, many were not,
and our object of interest was caught serendipitously in a MIPS scan
observation, or a MIPS photometry observation of another target.
Some were observed in more than one AOR.

For each observation, we extracted the individual basic calibrated
data (BCD) files  (individual flux density calibrated array images),
or enhanced BCD files, as appropriate, and pipeline-created
(``post-BCD") mosaics from the Spitzer Science Center (SSC) Spitzer
Heritage Archive (SHA) which were created using pipeline version
S18.13.  We investigated each mosaic for instrumental artifacts such
as jailbars for bright sources or dark latents from bright objects
observed prior to our observations. All of the mosaics were
sufficiently clear of artifacts at the position of our source of
interest, so we used the pipeline-produced mosaics for our 
targets. Our targets are bright, but none were saturated.

\subsection{Photometry}
\label{sec:photom}

For each of the final $\sim$260 MIPS-24 mosaics, we used the
APEX-1Frame package from MOPEX (Makovoz and Marleau 2005) to perform
point-response function (PRF) fitting to obtain photometry for each
object using the PRFs provided on the SSC website, again using the
standard SSC-recommended set of parameters.  At the same time, we also
used APEX to perform aperture photometry on the targets, as a check on
the flux densities we obtained from PRF fitting, to make sure that the
results were consistent.  For the aperture photometry, we used an
aperture radius of 13$\arcsec$ and an annulus of $20\arcsec$ to
$32\arcsec$, and the corresponding aperture correction of 1.17 from
the MIPS Instrument Handbook on the SSC website.  With these choices,
we find the PRF and aperture photometry yield consistent results, with
the mean difference being $<$PRF $-$ aperture$>$ = 0.030 $\pm$ 0.007
mag.  


In four cases, nearby objects are at risk of compromising the MIPS-24
photometry, which we now discuss.  HIP 9727 is a visual binary, but it
is also very bright. APEX erroneously identified two sources at the
location of the target (which is known to happen for very bright
objects such as this target); we confirmed using PSF subtraction
(module apex\_qa) that there are unlikely to be two sources cleanly
detected by APEX. In that case, the aperture photometry was used
instead of the PRF-fitting photometry. (We do not identify this object
as having an IR excess -- $K_s-[24]=0.04$, $[3.4]-[22]=0.01$.)


For two objects, HIP 105547 and HIP 99680, APEX identified more than
one source within $\sim5.5\arcsec$ of the desired target. Using PSF
subtraction (module apex\_qa), it is clear that both sources are
legitimately detected in both mosaics. The PRF-fitted flux density at
24 \mum\ is used for these two sources, and the WISE measurements are
tagged as unreliable, as WISE is unlikely to be unable to resolve each
of these pairs of objects (e.g., $\chi_{best}=\chi_{24}$; see below).
We note that HIP 105547 is reported in the literature as a binary, but
HIP 99680 is not. 


In the last case, HIP 32435, there are MIPS data in the SHA, but by
looking at the image, we determined that this object seemed to be
confused with a background galaxy. We dropped this object from our
sample entirely, as determining a flux density from the star alone
would be difficult (using MIPS or WISE).  


In a few observations of objects pulled out of the archive, where our
object of interest was caught serendipitously, sometimes the object
fell in a region with less-than-full coverage (HIP 19480, HIP 67103,
HIP 72603, HIP 78547). Since our objects are bright, even if there
were fewer MIPS frames than `normal' for most of the observation,
often adequate photometry could be obtained. We do not identify any of
these objects as having an IR excess.


In some of the observations pulled out of the archive, more than one
observation was obtained of our target (HIP 1427, HIP 3961, HIP 20491,
HIP 27633, HIP 66121). In these cases, we measured photometry
independently on the two (or three) pipeline mosaics (one is created
per AOR in the pipeline) and took a weighted average as the final best
value for the 24 \mum\ flux density.  Just one of these objects are
included in the objects we identify as having an IR excess: HIP 1427.


APEX provides the internal statistical uncertainties on the 24 \mum\
flux densities, and are $\sim$1\% or less. We have added a 2\%
systematic calibration uncertainty (Engelbracht et al.\ 2007, Gizis
2009, priv.~comm.) on top of this; this is added in quadrature to the
errors returned by APEX to obtain the errors we used.  We converted
the measured 24 \mum\ flux densities and uncertainties using the
zero-point magnitude of 7.17 Jy, from the SSC website (derived by
Rieke et al.\ 2008). 

The stars and their AORKEYs are displayed in
Table~\ref{fulltab}, along with the 24 \mum\ and the supporting
photometry obtained from the literature (see next section).

\subsection{Supporting data} 
\label{sec:sdata}

\subsubsection{Optical and NIR Data}

Ancillary optical data were primarily obtained from the merging of the
Hipparcos catalog with the literature.  The $B$ and $V$ photometry
comes from Tycho-2, where the Bessell (2000) formula was used to
convert $B_T-V_T$ to $(B-V)$ on the Johnson system.  The $JHK_s$
photometry is from the Two Micron All Sky Survey (2MASS; Skrutskie
\etal\ 2006).  Below, we will be using $K_s-$[24] as one indicator of
excess; we chose to use this rather than $V-[24]$ because of the
uniformity and consistency of values available via 2MASS for $K_s$.  
Rotational velocity measurements mostly come from Nordstrom et al.\
(2004), though some come from Holmberg \etal\ (2007) or Schroeder
\etal\ (2009). In our catalog, \vsini $\leq$ 3 km s$^{-1}$ should be
interpreted as an upper limit.  Operationally, we treat the upper
limits as detections because whether the rotational velocity is 3 km
s$^{-1}$ or $<$3 km s$^{-1}$ does not matter since they are all slow
rotators.

\subsubsection{Objects with very bright or incorrect $K_s$}

2MASS, as reported in the documentation (Cutri et al.\ 2003; Huchra et
al.\ 1994), does not acquire particularly reliable photometry due to
saturation for stars with $K_s < 3.5$, and stars with $K_s < 4.0$
often have large errors.  Quite a few stars in our sample (18) are in
this regime, and we now describe how we obtained more accurate $K_s$
estimates for these stars. The resulting $K_s$ magnitudes for these
stars, with their respective $K_s-[24]$  and [3.4]$-$[22] colors, can
be found in Table~\ref{brightk}. (Table~\ref{fulltab} also contains
the $K_s$ values we used.)  

The brightest object in our sample is HIP 22449, which has a $K_s$ in
2MASS reported as 1.6. Additional literature estimates could bring it
perhaps to $\sim$2 mag. Its brightness as measured in [24], [3.4], and
[22] are all $\sim$2 mag. Using the best possible $K_s$ we have, 
$K_s-[24]=-0.14$ and $[3.4]-[22]=0.22$. This object could have an IR
excess based on the WISE color, though it is so bright that it is
often used as a standard in the literature, and no one to date has
reported indications of a disk around it. We drop this object from
consideration in our sample because it is so bright that our ability
to obtain high-confidence estimates of its brightess is limited.


For the remaining bright stars, we estimate $K_s$ from the published
$V$ and $B-V$ photometry and a relation between $B-V$ and $V-K_s$
derived from high probability members of the Praesepe open cluster. 
Praesepe is near enough that its F stars have high quality optical
photometry, but far enough away that those same stars have good 2MASS
$K_s$ magnitudes.   The relation derived from the Praesepe stars is
$(V-K_s) = 2.219(B - V) + 0.07$.  The root-mean-square (RMS) of the
Praesepe stars about that relation is 0.059, which we take to be the
uncertainty in the derived $K_s$ (compared to stars with directly
measured $K_s$).

The known excess star HIP 61174 is the only star in this group with
a 22 or 24 \mum\ excess; it has $K_s-[24]=0.54$ and $[3.4]-[22]=0.67$.

HIP 8805 is not particularly bright, but is a visual binary in the
blue, slowly rotating group, with a separation of about 3.8$^{\prime
\prime}$.  The Hipparcos and Tycho-2 data suggest $V\sim$ 7.51 and
8.14 mags, and $B-V\sim$ 0.38 and 0.52 mags, for components A and B,
respectively.  The 2MASS point source catalog claims to provide
photometry for both components, but the listed values seem too bright
and too similar ($K_s=$ 6.227 and 6.280 mags, respectively).   We
removed this star from our analysis because of the unreliable near-IR
photometry, but still provide the data in Table~\ref{fulltab} for
reference. Using the data as provided in Table~\ref{fulltab}, this
star has one of the bluest $K_s-[24]$ values of $-$0.40 mag, lending
credence to our suspicion that something is wrong with the $K_s$
photometry. It has a very small WISE color, [3.4]$-$[22]=0.05 mag, so
we suspect that this object does not actually have an IR excess or a
disk. We have left it in the sample as a WISE-only source (e.g.,
$\chi_{best}=\chi_{22}$; see below), and have not identified it as
having a disk.


\subsubsection{Objects with large $K_s$ errors}

Four objects (HIP 9782, HIP 38423, HIP 66290, HIP 107339) have listed
$K_s$-band  uncertainties in the 2MASS PSC (Point Source Catalog) of
$>$9 mags.  The documentation for the PSC (Cutri et al.\ 2003)
indicate that these should not be interpreted as actual uncertainties,
but as flags to indicate that the pipeline was unable to accurately
estimate the uncertainty.  Because good estimates of the uncertainty
are critical (see below) for our determination of an IR excess, we now
address better values for these errors.

HIP 38423 is also very bright, with $K_s<4$. It has $J$ and $H$
errors of $\sim$0.3 mag. Because we redetermined a $K$ magnitude for
it above, it has a $K$ error of 0.06 mag. We do not identify an IR
excess in this object.

The other three sources are not particularly bright; HIP 9782 has
$K_s=6.74$, HIP 66290 has $K_s=5.63$, and HIP 107339 has $K_s=5.56$. 
The $J$ and $H$ uncertainties for all three of these stars are
``normal", with values of order 0.02 mag.  Using standard color-color
relations, we combined estimates of $J-K_s$ and $H-K_s$ (based on the
$B-V$ color) with the measured $J$ and $H$ magnitudes to predict a
$K_s$.  The predicted $K_s$ magnitudes are quite close to the observed
values, leading us to conclude that the published values are reliable,
despite the unusual uncertainty flags. We take the errors on the $K_s$
mag for these three stars as being 0.05 mag. None of these three
objects appear to have IR excesses. 

We note for completeness here that one source, HIP 61174, also known
as $\eta$ CrV, has relatively large 2MASS errors in all three bands;
errors are reported as $\sim$0.2-0.3 mags. This source is known to
have a disk. Since it has a bright $K_s$ mag above, it has an adjusted
$K_s$ and thus has adopted $K_s$ errors of 0.15 mag. 


\subsubsection{WISE data}

As mentioned above, we can use the WISE all-sky catalog to complement
our sample. We extracted WISE measurements from the WISE catalog for
each of our targets, matched by position. We took the  PSF-fitted
magnitudes and errors as reported by the catalog. It is important to
note that the 3.4 \mum\ (W1) detections for the entire sample are
saturated in the original data, and the values used here are obtained
via PSF fitting to the wings of the detection. The 22 \mum\ (W4)
points are, in contrast, not saturated. (The WISE Explanatory
Supplement states that the channels saturate at 8.1, 6.7, 3.8 and -0.4
mag for W1, W2, W3, and W4, respectively, but that fits to the
unsaturated wings of the PSF allow viable magnitudes to be obtained up
to 2.0, 1.5, -3.0 and -4 mag for the four channels, respectively.) 
For the majority of our sample, 255 stars, we have both [24] and [22]
measurements. The 22 \mum\ points are generally offset from the 24
\mum\ points; see the Appendix for more details.

Out of our original sample defined above, there were a handful of
objects that we removed from our proposal because we erroneously
thought that there were Spitzer observations in the archive.  We now
thus include 10 WISE-only objects in our sample (see
Table~\ref{fulltab}).  We consider the implications of including these
objects in our sample below. The blue-rapid sample has one WISE-only
object, blue-slow has one that we ultimately dropped from the final
sample (HIP 55897), red-rapid has six WISE-only objects, and red-slow
has two. 


For each of the WISE-only objects, we investigated the images at each
WISE band.  Most of them appeared to be relatively clean point
sources. HIP 32702 (in the red-rapid sample) seems to have somewhat
unusual structure in the 22 \mum\ PSF, and has $[3.4]-[22]=-0.13$; we
do not identify it as having an IR excess, and have left it in the
sample.  HIP 55897 (which would have been part of the blue-slow
sample) nominally has a [3.4]$-$[22]=0.29, but by inspection of the W4
image, we consider it unlikely to have a real excess. We include it in
Table~\ref{fulltab} for completeness, but drop it from the rest of the
analysis. 


There is one object that seems, by comparison to the $K_s-[24]$ values
as well as comparisons within WISE (e.g., [3.4]$-$[12]) to have either
the 3.4 or 22 \mum\ measurement corrupted. HIP 3505 has $K_s-[24]=0$
and $[3.4]-[22]=0.90$. We suspect that this object does not really
have a disk, and regard its WISE data as unreliable. 


\subsubsection{Other IR bands}

Six of the stars considered here, such as HIP 61174 = $\eta$ CrV and
HIP 76829 = HD 139664, are already identified in the literature as
having IR excesses due to a circumstellar disk, though often using (or
including) wavelengths longer than 24 \mum. For the purposes of our
discussion here, we did not include information from wavelengths
longer than 24 \mum. We also did {\bf not} include information, if
available, from 4-20 \mum. For the purposes of our discussion here, we
wished to not introduce biases by including information for some
objects (likely preferentially those with disks) obtained at other
bands. We establish the presence or absence of an IR excess below
using 22-24 \mum\ alone. (See Table~\ref{tab:disks} for a complete
list of literature disks identified using any wavelength.)

\subsubsection{Final sample sizes}

In the end, there are 263 stars in the final sample, having dropped
two above, one with MIPS data and one WISE-only object. There are  49
stars in the blue-slow sample, 59 stars in the blue-rapid sample, 78
stars in the red-slow sample, and 77 stars in the red-rapid sample. We
present new MIPS measurements for 188 objects, measurements from MIPS
data already in the archive for 67 objects (one of which we  dropped
out of the sample), and WISE-only data for 10 objects (one of which we
dropped out of the sample).

\section{Identification of Infrared Excess Objects}
\label{sec:disks}

\subsection{Selecting the objects with excesses}

In order to identify those objects that have an apparent IR excess and
thus those that we will infer to have a circumstellar disk, we want to
compare shorter-wavelength and longer-wavelength bands. We need a band
that samples photosphere and a band that samples where the debris disk
is expected to be bright. Preferably the two bands should be both on
the Rayleigh-Jeans (RJ) side of the SED for these F stars, and be
measurements that are as uniform (homogeneous) and well-calibrated as
possible, such that the scatter in the measurements (e.g., for those
objects without disks) is minimized.  

We have $V$ magnitudes for all of the stars in our sample, and the $V$
magnitude is likely to be sampling photosphere in these stars.
However, the $V$ measurements come from a variety of sources (so there
is a lot of scatter in these values introduced by their various
different origins), and $V$ band for this sample of F stars is near
the peak of the SED, so any color involving $V$ (e.g., $V-K_s$ or
$V-[24]$) will not have an expectation value of zero.

We have $K_s$ for all of the stars in our sample from 2MASS, and as
such there is minimal scatter introduced from variations in
calibration, data reduction systematics, etc. (except for the
brightest stars, as discussed above). For these stars, which are not
likely to have substantial inner disks, $K_s$ is likely to be sampling
photosphere. We checked the validity of all the $K_s$ magnitudes as
discussed above.  We examined the $JHK_s$ color-color diagram for our
sample for any evidence of excesses at $K_s$ and found none. Moreover,
$K_s$ is on the RJ side of the SED, such that, e.g., $K_s-[24]$ is
expected to be zero and not vary over the (relatively small) range of
types included here. 

We have 24 \mum\ measurements for 97\% of our sample. These data have
all been re-reduced here so as to minimize data reduction systematics.
MIPS data are known to be sensitive and have a stable PSF, such that
the internal statstical error for the photometry is quite low. However
the PSF, at 6$\arcsec$, can include background (or foreground) objects
close to the target. We selected our sample so as to be in relatively
uncrowded fields so as to minimize this; see also discussion below
about contamination.

We also have WISE measurements at 3.4 and 22 \mum\ for all of our
sample. These data are all uniform and internally consistently
calibrated. The 3.4 \mum\ band, while longer than the 2.2 \mum\ $K_s$
band, is likely to still sample photosphere, and the 22 \mum\ band
should sample dust at a comparable distance from the star as the 24
\mum\ band. The spatial resolution (at 12$\arcsec$) is poorer for WISE
(than for MIPS), but as these fields are relatively uncrowded, this
should have minimal impact. The data were obtained for the same
objects nearly simultaneously, which minimizes any intrinsic
variations in these stars at these bands.

So, which indicator should we use for identification of IR excess
sources? As discussed in the Appendix, the intermingling of $K_s$,
[24], and WISE is somewhat problematic. However, $K_s-[24]$ and
[3.4]$-$[22] are {\em two entirely independent measures} of IR
excess.  Figures \ref{fig:kk24} and \ref{fig:w1w4} show the $K_s$ vs.\
$K_s-[24]$ and [3.4] vs.\ [3.4]$-$[22] plots for the sample,
respectively.  As can be seen in the plots, these distributions are
similar in that the bulk of the distribution is scattered about 0 and
there are outliers to the red (right) that we infer to be the disk
candidates. The extremely red objects in Fig.~\ref{fig:kk24} and
\ref{fig:w1w4} are the same objects. In Fig.~\ref{fig:kk24}, the
distribution is relatively tightly distributed about the $K_s-[24]$=0
line, though it is slightly offset to the blue such that the most
likely value in the distribution (the mode of the distribution) is
actually $K_s-[24]=-0.03$ -- this is most likely primarily a result of
some combination of our particular reduction of the MIPS data, plus
uncertainties in absolute calibration between the 24 $\mu$m and $K_s$
systems. The distribution develops more scatter at the brighter end,
where the $K_s$ values are not as reliable (as per discussion above),
and there are fewer of them.   In Fig.~\ref{fig:w1w4}, [3.4]$-$[22] is
zero for most of the ensemble, and the scatter about the
[3.4]$-$[22]=0 distribution is somewhat larger than that for
$K_s-[24]$, but better-centered on 0, and does not appear to disperse
quite so obviously at the brighter end. 

We investigated seven options for determining where to set the limit
between disk candidate and photosphere candidate at 22 or 24 \mum. We
now discuss each of these in turn; the last option is the one we
selected.

{\bf (1)} We looked at simply setting a limit of +0.1 mag in the
$K_s-[24]$ color or +0.2 mag in the [3.4]$-$[22] color; there is a gap
at [3.4]$-$[22]=0.2 that is tempting to use. However, there is no such
gap in the $K_s-[24]$ distribution, and in theory, we are more
sensitive to weaker disks using [24] rather than [22].

{\bf (2)} We tested a method by Sierchio et al.\ (2010), which for a
given star uses its $B-V$ color to predict its $H-K$ color from the
Kenyon \& Hartman (1995) table.  From the observed $H$ and predicted
$H-K$ color, an estimated $K$ magnitude was calculated.  This
estimated $K$ magnitude was then averaged with the observed $K_s$
magnitude to produce what Sierchio et al.\ (2010) found to be a better
estimate of $K$ for their sample.  This method, for our data sample,
did not give a significantly lower scatter in our observed data, so we
used the observed $K_s$ magnitudes for this study, except where noted
in \S\ref{sec:sdata}.  

{\bf (3)} We tested a method suggested by Urban \etal\ (2012) which
uses $V-K_s$ to predict the $K_s-[24]$ values expected, but for the
narrow range of F stars considered here, this is equivalent to
assuming $K_s-[24]$ should be 0 (or indistinguishably close to it), so
it yields identical results to one of the other methods considered
here.

{\bf (4)} We empirically assessed the scatter in the $K_s-[24]$ and
[3.4]$-$[22] distributions as a function of brightess via computing a
running mean and standard deviation ($\sigma$), and attempted to set a
3-$\sigma$ limit based on that scatter. The scatter in our $K_s-[24]$
data is consistent with (or better than) other similar studies in the
literature.  The scatter indicates the overall empirical uncertainty
in the calculated $K_s-[24]$ and [3.4]$-$[22] distributions, and gets
larger at fainter magnitudes; however, it does not take into account
the uncertainties on the individual measurements, and it becomes more
difficult to calculate the scatter at the bright end, due to
decreasing numbers of objects.

{\bf (5) and (6)} We incorporated the errors on the measurements by
calculating \begin{equation} \chi_{24} = \frac{(K_s-[24])_{\rm
observed} - (K_s-[24])_{\rm predicted}}{\sigma_{(K_s-[24])}}
\end{equation} as per, e.g., Trilling \etal\ (2008).    For
$\chi_{24}$, the expected (predicted) $K_s-[24]$ for our data
reduction is evidently slightly blue at $-0.03$, since that is the
most likely value found in the distribution of $K_s-[24]$ values. As a
completely different assessment (which we take to be disk option 6),
we similarly calculated $\chi_{22}$; we took the expected value of
[3.4]$-$[22] to be 0. The limit for significant IR excess is usually
taken to be $\chi_{x}=3$. When selecting IR excess objects using just
$\chi_{24}>3$, 21 objects are selected; using just $\chi_{22}$, 11
objects are selected. These objects are not quite the same between the
two selections, due to uncertainties in the individual points, e.g.,
cases like HIP 8805 discussed above, where $K_s$ is likely wrong.
There are 9 objects for which both $\chi_{24}$ and $\chi_{22}$ are
$>$3: HIP 18187, HIP 18859, HIP 20693, HIP 24947, HIP 25183, HIP
61174, HIP 88399 HIP 99273, HIP 100800. There are, however, many
objects with ``borderline'' $\chi$ values, more so for $\chi_{24}$
than $\chi_{22}$ -- for these objects, $\chi_{24}$ is near but not
quite 3 (more on these objects below).

Essentially all of these methods select the same 7 stars with the most
extreme values of $K_s-[24]$ and [3.4]$-$[22], and these are likely
the most secure disk candidates (HIP 18187, HIP 18859, HIP 20693, HIP
25183, HIP 61174, HIP 88399, HIP 99273). These objects all have
$K_s-[24] \geq 0.265$ and $[3.4]-[22] \geq 0.295$ -- though we note
there is one more IR excess object (HIP 24947) with $K_s-[24]=0.265$ but
$[3.4]-[22]=0.244$, so not quite as extreme in $[3.4]-[22]$, though
clearly identified as having an IR excess.  

There are many objects selected by only one or two methods out of the
several we tested.  There are some objects that have obviously
incorrect values in one color but not the other, rendering methods
using only the `wrong' color less than optimal. For example, HIP
8805, discussed above as likely having an incorrect $K_s$, has
$K_s-[24]=-0.40$ but [3.4]$-$[22]=0.05, suggesting that indeed, its
$K_s$ is incorrect, and that it doesn't have a detectable IR excess.
It can also happen that one or more of the WISE bands are likely
incorrect; as discussed above, HIP 99680 has [3.4]$-$[22]=$-0.37$, the
bluest of the ensemble, and $K_s-[24]=0.14$, so it is likely that
either [3.4] or [22] is not correct for this object.   However, the
overwhelming majority of the sample has $K_s-[24]$ and [3.4]$-$[22]
measurements that are consistent with each other (and therefore
$\chi_{24}$ and  $\chi_{22}$ values that are consistent with each
other).  There are objects that show up as marginally
significant in both bands independently, e.g.
$\chi_{22}\sim\chi_{24}\sim2.8$.

We examined the the distributions of $\chi_{22}$ and $\chi_{24}$; see
Figure~\ref{fig:chihisto}. The histograms are generally well behaved,
in that there is a strong peak near 0 for both. The outliers on the
blue ($\chi_{x}<0$) side can, in both cases, be attributed (as for HIP
8805 or HIP 99680) to bad values in one of the relevant bands. Both
distributions have a long tail to the red ($\chi>0$), as expected. The
histogram for $\chi_{22}$ is nicely Gaussian, and $\chi_{22}=3$ seems
visually to be the right cutoff given the distribution; there is a gap
near $\chi_{22}\sim3$ corresponding essentially to the gap seen near
[3.4]$-$[22]=0.2. The best-fit Gaussian to this distribution suggests
that a 3-$\sigma$ cutoff would be $\chi_{22}\sim$2.4 rather than 3. 
The histogram for $\chi_{24}$ is not quite as clearly Gaussian, with a
more prominent asymmetry to the red. While there is also a visible gap
near $\chi_{24}\sim3$, and a Gaussian fit to the distribution suggests
that a 3-$\sigma$ cutoff would be $\chi_{24}\sim$2.5 rather than 3,
the $\chi_{24}$ distribution deviates from a Gaussian with a red
``bump'' near  $\chi_{24}=2$ such that there are $\sim$15 objects
right at that boundary. 

{\bf (7)} We would like to take into account information from all four
bands ($K_s$, [3.4], [22], and [24]), compensating for missing or
obviously incorrect values, and make a viable decision on what objects
should or should not be disk candidates. Since $K_s-[24]$ and
[3.4]$-$[22] are two entirely different and independent measures of IR
excess, we wished to combine them.  We calculated $\chi_{22}$ and
$\chi_{24}$ for each and then took all objects for which
\begin{equation} \chi_{\rm best} =
\frac{\chi_{22}\sigma_{(K_s-[24])}+\chi_{24}\sigma_{([3.4]-[22])}}{\sqrt{
\sigma_{([3.4]-[22])}^{2} + \sigma_{(K_s-[24])}^{2}}} \geq
3\end{equation} as IR excess candidates. In some cases, where data are
missing or obviously incorrect, the $\chi_{\rm best}$ was taken to be
either $\chi_{22}$ or $\chi_{24}$, as appropriate, and these are noted
in Table~\ref{tab:disks}.  This merging of the two different values of
$\chi$ (where available) ameliorates inaccurate measurements in any of
the four bands (particularly those for which, e.g., one band has a
large uncertainty), takes advantage of all the information we have,
and incorporates the idea that the significance of the excess should
be inexorably tied to the size of the uncertainty on the measurement. 
It includes as disks those things that are borderline in both (e.g.,
$\chi$=2.5 in both independent measures, or 2.97 in one and 2.5 in the
other), and does not include those things that are borderline in one
but not the other (e.g., 2.97 in one and 0.8 in the other). There are
13 objects in our final IR excess selection with $\chi_{22}<3$. There
is one object with $\chi_{24}<3$, and it also happens to have
$\chi_{22}<3$ (HIP 47198). 

The distribution of $\chi_{\rm best}$ is shown in
Figure~\ref{fig:chihisto}, and it is the least Gaussian of the three
distributions in that Figure.  The definition of $\chi_{\rm best}$ is
consistent with this effect. The best-fit Gaussian is consistent with
a cutoff of $\chi_{\rm best}=3$ for a significant excess.

All values of $\chi$ are provided in Table~\ref{fulltab} so that
readers can make their own judgements as to the strength and
reliability of the excess. Note that $K_s$, [24], [3.4], [22],
$K_s-[24]$, and [3.4]$-$[22] are given (with errors) to 0.001 mag --
more than is necessarily warranted given the precision of the results,
but this aids in reproduction of our values of $\chi$.  We are taking
$\chi_{\rm best} \geq$3 as our IR excess candidates for the remainder
of this paper.  The objects indicated as having an IR excess in all
the figures here are those selected via this mechanism.

\subsection{Limits on contamination}

Empirically, within the $\sim$300 objects we examined in the course of
this work, one object (HIP 32435) was discarded as confused with a
background galaxy.  We now consider those cases in which we would not
be able to resolve and thus identify a contaminant, whether a
background galaxy or an asteroid.

As in Stauffer et al.\ (2005), we determined the expected rate at
which background active galaxies projected onto the line of sight of
our target stars would produce false-positive apparent infrared
excesses at 24 \mum.  We used the 24 \mum\ source counts in Papovich
et al.\ (2004) to calculate the probability of a chance superposition
of a ULIRG and of our target stars.  Because the contamination rate
increases towards lower flux levels, as a conservative measure we only
calculated the contamination rate for our faintest sources.  There are
98 stars in our sample with $K_s >$ 6. Out of those 98, 92 have a 24
\mum\ flux measurement; their mean 24 \mum\ flux density is $\sim$22
mJy.  Our smallest excess limit of $\sim$8\% then implies that a
contaminating ULIRG would need a flux density greater than $\sim$1.8
mJy to cause us to misidentify a bare photosphere as an IR excess
object. Using data from Fig.~2 of Papovich et al.\ (2004), there are
$\sim10^6$ obj sr$^{-1}$ or $\sim2.3\times10^{-5}$ obj arcsec$^{-2}$
with 24 \mum\ flux density greater than this.  Combining this estimate
with our 13$^{\prime\prime}$ aperture radius yields a predicted
contamination rate of 0.012 for each of our faint target stars, or a
1.2\% chance that one of our faint stars will have a false excess
detection due to ULIRG contamination.  This rate will be even lower
since we are using PSF fitting photometry, not aperture photometry,
for most of our MIPS sources. WISE data, though it has a larger PSF,
is also PSF-fitted photometry, and is overall less sensitive than
MIPS; the net rate of contamination in the WISE data is likely to be
comparable.


Asteroids are not a concern for this project.  Taking the Taurus
Spitzer Survey (Padgett et al.\ 2012 in prep; Rebull et al.\ 2010) as
a worst-case scenario because the survey is taken in the ecliptic
plane, we would expect 0.0032 asteroids on a given aperture, even in
the ecliptic plane.  When taken 300 times, to approximate our sample,
we expect to find an asteroid contaminating our target 0.96\% of the
time.  Since our targets are all over the sky and not all in the
ecliptic plane, the real contamination rate is even lower.  While the
WISE PSF is larger, the data acquisition scheme for WISE is such that
moving targets can be removed, and so it is very unlikely that
anything remaining in the WISE catalog are asteroids.

To investigate the chances of randomly finding stars with excesses
(due to real or contaminated sources), we conducted Monte Carlo
experiments.  For groups of 50 stars (to approximate each of our
color/rotation groups), we drew a random set of stars from a master
pot of simulated stars with a known disk fraction.  For a ``true" disk
fraction of 5\%, there is a $<10$\% chance of finding 5 or more disks,
as for the red rapid sample.  For a ``true" disk fraction of 2.5\%,
there is a $<0.6$\% chance of finding 5 or more disks.  This is
consistent with expectations from binomial statistics (11\% chance of
5 or more disks for P=0.05 and  0.8\% chance for 5 or more disks for
P=0.025).

\subsection{WISE additions}

We have added to our sample some objects with only WISE data, no MIPS
data.  In order to investigate whether this might be implicitly
biasing our results, we looked more closely at the distributions of
$\chi_{22}$ and $\chi_{24}$. 

For the majority of our sample, $\chi_{22}$ and  $\chi_{24}$ are
consistent with each other. If there is a large $K_s-[24]$, then there
is also a large [3.4]$-$[22] and vice versa. In those borderline cases
where the $\chi$ from either $K_s-[24]$ or [3.4]$-$[22] is near 3 but
less than 3, the additional information obtained from the additional
measurements can move the object over into the regime of believable IR
excesses.  For this sample, the distribution of $\chi_{22}$ seems to
be somewhat better-behaved in that the distribution is closer to a
Gaussian. Out of the 13 objects that have $\chi_{22}<3$
and/or $\chi_{24}<3$ but  $\chi_{\rm best}>3$, all also have 
$\chi_{22}<\chi_{24}$. If we had only the WISE data for these stars,
these borderline objects would not have been selected as IR excess
objects.  Among the sample of objects selected to have IR excesses,
the lowest value of $\chi_{22}$ for any object with well-behaved [3.4]
and [22] is $\sim$1.2.

There are 9 WISE-only sources in the final sample scattered fairly
evenly through the distributions. None of the WISE-only sources have
$\chi_{22}>3$.   There are four sources out of those 9 that have
$\chi_{22}>1.2$, so these, if we had supporting MIPS data, might have
small IR excesses, e.g., weak disks.  This is likely within the errors
we consider below.


\subsection{Binaries}

Our sample selection mitigated against equal-mass or near-equal-mass
binaries, but many are in fact known (apparent) binaries, and our
knowledge of this information may be incomplete. 

Low-mass companions could make an apparent IR excess. A simple
comparison of blackbody models at the appropriate $T_{\rm eff}$
suggests that an M star companion to our F stars could be contributing
between $\sim$25\% and 35\% of the flux at 24 \mum.  Among our
candidates for IR excess, the distribution of fractional measured
excesses (meaning (measured-predicted)/predicted), is broad, but there
is a pileup of values at the 15-20\% level. If the more conservative
criterion of $\chi_{24}>3$ is adopted, there is still a pileup of
fractional excess at 24 \mum\ at the 15-20\% level; if one adopts
$\chi_{22}>3$, the pileup occurs at $\sim$30\%.  

However, if there is an M star companion, $K_s-[24]$ would not be zero
for that photosphere (Gautier \etal\ 2007). Taking $K_s-[24]\sim$0.3
as typical, if there is an unresolved M star that makes up 20\% of the
total flux, then the net photosphere might be $\sim$5-10\% brighter in
$K_s-[24]$.  Moreover, this issue would only matter if there was an
apparent binary with a separation of a few arcseconds such that $K_s$,
the highest spatial resolution observation of the four bands, resolved
a pair of stars, but MIPS did not. In this case, neither WISE band
would be able to resolve the stars either, and the [3.4]$-$[22] would
be less affected than the $K_s-[24]$. It is unlikely that there are
many cases like this in our sample.

When obtaining the measured \vsini, some binaries were identified --
59 objects (23\%) out of the entire set. These are indicated in
Table~\ref{tab:disks}. However, only two of them are identified as
having an excess, HIP 25183 and HIP 61621. It does not appear that
binaries are a significant problem for our dataset.


\subsection{Net Disk (Excess) Frequency}

Having established our criteria for selecting significant IR excesses
above, we find a net $\sim$8\% excess rate; 22 out of 263 stars in our
sample have significant infrared excesses at 22-24 \mum.  
Figures~\ref{KvK24_full} and \ref{fig:w1w4} identify those stars with
excesses; Figure~\ref{KvK24_full_indiv} plots $K_s$ vs. $K_s-[24]$ for
each sub-sample individually, with excesses marked.  

Overall, three stars in our sample (HIP 61174, HIP 88399, HIP 99273)
have $K_s-[24]>0.5$ (and $[3.4]-[22]>0.6$), and are the most
unambiguous disks. All three of them are rapid rotators, and
previously-identified disks (see Table~\ref{tab:disks}). HIP 61174 is
$\eta$ CrV, first noted as having a disk by Stencel \& Backman (1991),
and is the brightest disk of the sample. HIP 88399 is HD 164249, part
of the Beta Pic Moving Group, and identified by several authors as
having a disk (Mo\'or \etal\ 2011; Rebull \etal\ 2008; Rhee \etal\
2007).  HIP 99273, the largest $K_s-[24]$ and $[3.4]-[22]$ in the
sample, is HD 191089, identified as having a disk by Rhee \etal\
(2007). 

Out of the next five stars in our sample with the next largest
$K_s-[24]$ excesses (HIP 18187, HIP 18859, HIP 20693, HIP 24947, HIP
25183), two have  previously been noted as having a disk. HIP 18859 is
HD 2545, and was identified as a disk by Rhee \etal\ (2007). HIP 24947
is AS Col, which was identified as a disk by Zuckerman \etal\ (2011).
The rest are newly identified excess stars. They have $K_s-[24]$ from
0.27-0.3, and $[3.4]-[22]$ from 0.24 to 0.46. Another star, HIP
100800, is next in ordering by 24 \mum\ excess, with $K_s-[24]=0.18$
but $[3.4]-[22]=0.28$; this is another newly identified disk. 

The remaining 13 stars we identify as having an IR excess (see
Tables~\ref{fulltab} and \ref{tab:disks})  are all more moderate
excesses, with $K_s-[24]$ between 0.03 and 0.13 mag, and [3.4]$-$[22]
between 0.07 and 0.18 mag -- with two of the objects having possibly
questionable [3.4]$-$[22]$<$0. Four of these lower $K_s-[24]$ objects
are previously identified disks in the literature -- HIP 108809 (Rhee
\etal\ 2007), HIP 107947 (Zuckerman \etal\ 2011), HIP 49809 (Trilling
\etal\ 2007), and HIP 114948 (Beichman \etal\ 2006).  There is only
one of these lower-excess objects in the blue-rapid sample; the rest
are scattered amongst the other three sub-samples.

How does our overall excess frequency compare to previous
determinations for similar target populations?   Perhaps the best
comparison is to the Formation and Evolution of Planetary Systems
(FEPS) sample (Meyer  et al.\ 2006, Carpenter et al.\ 2008, 2009). 
FEPS obtained Spitzer data using all the instruments for a sample of
314 F, G and K dwarfs.   Because the FEPS program was designed to
determine how debris disks evolve with time, their sample was
significantly weighted to young ages where debris disks are {\em a
priori} more likely to be detectable; in particular about 20\% of
their stars are younger than 20 Myr, whereas few to none of our sample
are likely to be that young. Our stars are expected to be field stars,
and $\sim$60\% of the FEPS sample is also field stars. The  F stars
make up only about 13\% of the entire FEPS sample, and about 15\% of
the sample $>$20 Myr. Excluding the age $<$20 Myr stars from their
sample (but retaining stars of all spectral types), and using their
flux density ratio of 24 to 8 \mum\ as a proxy for the excess above
photospheric flux, 21 of 232 of their stars have 24 \mum\ excesses
exceeding 10\% above the photosphere. This yields an excess (disk)
frequency of about 9\%, slightly more than what we obtain (8\%).  
They note that the excess frequency drops at older ages, to $\sim$3\%
for age $>$ 300 Myr.  Our sample of stars is more likely to be closer
in age to 300 Myr than theirs, on average. On the other hand, Meyer et
al.~(2008) note that the cluster stars show a higher excess fraction
(at 24 \mum) than a comparable field sample of a given age; our stars
are likely to be field stars. Our excess frequency of $\sim$8\% is
consistent with their measurement.

Another recent sample of F stars, Mo\'or et al.~(2011), finds 27 disks
out of 82 F stars, or 33\%. However, as discussed in that paper, the
sample was deliberately selected so as to be biased towards likely
harboring a disk. Thus, that sample cannot be directly compared to our
sample.

\section{Discussion and Conclusions}
\label{sec:results}

The disk (excess) fractions are listed explicitly in
Table~\ref{excess}.  The blue-rapid sample has the lowest disk
fraction at $\sim$3\%, and the red-slow has the next lowest disk
fraction at $\sim$6\%. The blue-slow sample has a disk fraction of
$\sim$8\%, and the red-rapid sample has the highest disk fraction at
$\sim$14\%. The errors on these disk fractions can be calculated using
the binomial distribution, as per Burgasser \etal\ (2003);
Table~\ref{excess} lists these formal errors, and Figure~\ref{fig:df}
represents that information graphically. 

Among the entire sample, we find the highest fraction of IR excesses
(and inferred disks) in the red-rapid rotators ($\sim$14\%).  There
are $\sim$8\% disks in the sample opposite in  properties, the blue
slowly rotating sample.  The lowest disk fraction is the blue rapid
sample ($\sim$3\%), which is consistent with the blue slow and red
slow samples and is perhaps distinct in disk fraction from the red
rapid sample ($\sim$14\%).   The red-rapid sample may have
significantly more disks than the blue-rapid sample, but both blue
samples are consistent with each other, and both red samples are
(barely) consistent with each other. Both slow samples are consistent
with each other, and the two rapid samples may be significantly
different from each other.

Taken at face value, these disk fractions are not what we expected,
{\em a priori}, to find.  If winds clear disks, and if early F stars
do not have winds, then we would have expected a high disk fraction
for both of the blue samples (total of 6/108=6\%, error range 4-9\%)
compared to the red samples (total of 16/155=10\%, error range
8-13\%).   We do not see that; they are even consistent with each
other, taking both extremes of the errors.   The division
between red and blue stars corresponding to a type of F5 is not
particularly subject to debate. The nature of the winds on either side
of this boundary almost certainly is different. However, we cannot
discern any significant difference in the disk fractions between the
red and blue samples. 

The division we used between rapid and slow rotators is 15 km
s$^{-1}$. As discussed above, this was selected based on where the
X-ray flux saturates for G stars, and the sub-samples were constructed
to be approximately equally sized, given this cutoff (136 rapidly
rotating objects, and 127 slowly rotating objects). Their overall disk
fractions are 13/136=10\% (error 8-13\%) and 9/127=7\% (error 5-10\%).  
Figure~\ref{vsini} shows a \vsini\ vs.\ $K_s-[24]$ plot for our whole
sample, with our rapid/slow division indicated in the plot. If one
instead were to use a division of order 30 km s$^{-1}$, however, this
plot appears to show a low excess frequency for rapid rotators and a
relatively high excess frequency for slow rotators.  However, the
correlation is illusory, in large part because of the distortion of
the sub-sample sizes (53 objects with \vsini\ $>$ 30 km s$^{-1}$ and
210 objects with \vsini\ $\leq$ 30 km s$^{-1}$).  Nearly 65\% of the
53 stars with \vsini $>$ 30 km s$^{-1}$ are early F stars.  Were we to
use 30 km s$^{-1}$ instead as the cutoff between rapid and slow, the
disk fractions would be 2/34 (6\%, blue-rapid), 1/19 (5\%, red-rapid),
4/74 (5\%,blue-slow), 15/136 (11\%, red-slow).  The subsamples are so
different in size that comparison of them is very difficult, and these
disk fractions are identical to what we have already determined for
these samples within small number statistics. We will retain our
rapid/slow cutoff of 15 km s$^{-1}$ for the remainder of the paper.

The overall disk fraction we obtain here is quite consistent with the
disk fraction (at 24 \mum) obtained elsewhere. A significant
complication, however, within our sample and with comparison to other
samples, is the ages of the stars, which we now discuss.

Slow rotators with disks are certainly in our sample. A possible
explanation for the existence of slow rotators with excesses is
long-lived pre-main-sequence circumstellar disks.  A number of
theoretical models predict a strong correlation between rotation and
primordial disk lifetime (via ``disk locking"), whereby the angular
rotational velocities of PMS stars are magnetically locked to the
Keplerian velocity of the inner part of the star's disk as long as
accretion continues (e.g., K\"{o}nigl 1991, Shu et al.\ 2000).  Such
correlations are observed through mid-IR excesses (taken to arise from
primordial circumstellar dust) that are larger around slowly rotating
PMS stars (e.g., Rebull et al.\ 2006, Kundurthy et al.\ 2006).  These
models predict that stars with long-lived accretion disks will arrive
on the main sequence as slow rotators and those with short-lived
accretion disks will arrive on the main sequence as rapid rotators. 
If one assumes that stars with long-lived accretion disks are more
likely to form well-populated planetesimal belts, and therefore to
have detectable debris disks at later ages, then there should be an
anti-correlation between \vsini and 24 \mum\ excess for young low mass
main sequence stars.  However, whether or not we see this correlation
in our data set is certainly intimately tied to the issue of ages in
these stars.


While Figure~\ref{MvBV} demonstrates that there is no large,
systematic, and obvious age difference between our rapid and slow
rotators, it is almost certainly true that our red, rapidly rotating
sample is younger -- on average -- than our red, slowly rotating
sample.  This follows simply from the fact that stars in the red
sample ($(B-V) >$ 0.43) are expected to have dynamo-driven solar-type
winds and should thus spin down as they age. Slow rotators could
either be stars (of any age) with long-lived PMS disks that arrived on
the ZAMS as slow rotators or older stars that arrived on the MS as
rapid rotators but subsequently lost angular momentum due to winds. 
Rapid rotators must necessarily be relatively young.  Comparison to
the rotational velocities of stars in open clusters in this color
range -- Pleiades, age $\sim$100 Myr (Queloz et al.\ 1998); Hyades,
age $\sim$ 650 Myr (Mermilliod et al.\ 2009, Figure 4); and M67, age
$\sim$4 Gyr (Melo et al.\ 2001) -- suggests that the red-rapid sample,
and in particular the stars with excesses, probably have ages in the
range $<$100 Myr to $\sim$ 1 Gyr. (In fact, the sole blue rapid
rotator with a disk is also consistent with this younger age range,
but with only one disk, the significance of this comparison is
lessened.) See Figure~\ref{vsiniandpleiades} for the comparison to the
Pleiades.   This suggests that the red rapid rotators are on average
somewhat younger than the likely average age for the red slow
rotators, but not by a huge amount.  A younger age for the red rapid
sample would help explain their higher 24 $\mu$m excess frequency.  We
note that all of the stars with excesses have isochronal age estimates
provided in Holmberg et al.\ (2009).  The ages are all between 1.2 and
4.8 Gyr; the red-rapid sample makes up most of the disks, but is not
distinctly younger than the rest of the objects.  These ages seem at
variance to the ages inferred from their rotational velocities. For
these objects, Holmberg adopts metal poor metallicities ($<$[Fe/H]$>
\approx -$0.2), which may or may not be appropriate.   Spectroscopic
metallicity determinations for these stars would help considerably in
the interpretation of their ages and hence the origin of their disks.

Nevertheless, the mere existence of significant debris disk excesses
around the 11 members of the red-rapid sample is significant.
Either the red-rapid sample includes a set of stars that are
significantly younger than the rest of our sample (and the high disk
frequency simply reflects the fact that younger stars more frequently
have debris disks) or the red-rapid sample includes a set of stars of
``normal'' age but unusual dust content. Better ages for these stars
would address the first possibility; X-ray fluxes for this set of
stars could help constrain whether they have unusually weak winds (for
example) by use of the Wood \etal\ (2000) formula relating X-ray flux
and mass loss rate for low-mass stars.   Unfortunately, no X-ray
fluxes are available for the red-rapid rotators with IR
excesses, hence we cannot estimate their mass loss rates using the
Wood et al.\ (2000) formula.  X-ray data for these stars, and better
age estimates (perhaps obtainable by combining Gaia parallaxes and
improved [Fe/H] estimates) could place interesting constraints on the
dust production rates for these stars.

Sierchio et al.\ (2010), Stauffer et al.\ (2007), and Rebull et al.\
(2008) have all previously noted a possible correlation between 24
$\mu$m excess and rotation, with excesses being rare or absent amongst
rapid rotators.   The samples of stars used in these studies have
included F, G, and K dwarfs, and thus are on average significantly
later type than our sample.  Even if the correlation with rotation in
those samples is real (the statistical significance was not high), it
does not necessarily contradict our result.  The wind mass loss rates
from G and K dwarfs are likely much higher than for F dwarfs; the
comparison is made difficult because direct measurements are not
possible, and inferences from rotational velocities depend on the
unknown degree to which the spindown affects just the outer convective
envelope or also the radiative core. Moreover, the relative strengths
of wind and radiation pressure are important and F stars have higher
luminosities than later-type stars (the ratio of mass loss to total
luminosity is lower).  If the G and K dwarf mass loss rates are much
higher, they could successfully scour their disks of dust and produce
the apparent correlation, while the weaker winds of F dwarfs still
allow debris dust.

One of our original goals was to determine if F dwarfs have a
comparatively high debris disk frequency. By comparison to the FEPS GK
stars of similar ages, our F stars have a comparable disk frequency.
Another of our original goals was to look for evidence (via debris
disk frequency) for development of winds at the F5 boundary where
solar-type winds are expected to develop, and we do not see strong 
evidence of this.  A larger sample of F stars would presumably reveal
more disks, and thus enable better restrictions on the observed disk
fraction.  Observations in other infrared wavelength bands would also
enable us to see if our 24 \mum\ excesses extend to longer or shorter
wavelengths.  A wider range of observations would allow for more
conclusive evidence in understanding what is going on in these stars. 
It would also be of interest to compare our infrared observations with
X-ray observations to see if there is any indication of strong
activity, which would help to argue for or against the theory of Chen
et al.\ (2005).

\acknowledgments

The authors wish to acknowledge truly helpful comments from the
anonymous referee, and helpful conversations with Sidney Wolff.

This work is based in part on observations made with the Spitzer Space
Telescope, which is operated by the Jet Propulsion Laboratory,
California Institute of Technology under a contract with NASA. Support
for this work was provided by NASA through an award issued by
JPL/Caltech. The research described in this paper was partially
carried out at the Jet Propulsion Laboratory, California Institute of
Technology, under contract with the National Aeronautics and Space
Administration.    This research has made use of what was then called
the NASA/IPAC/NExScI Star and Exoplanet  Database, which was operated
by the Jet Propulsion Laboratory, California  Institute of Technology,
under contract with the National Aeronautics and  Space
Administration. 

This research has made use of NASA's Astrophysics Data System (ADS)
Abstract Service, and of the SIMBAD database, operated at CDS,
Strasbourg, France.  This research has made use of data products from
the Two Micron All-Sky Survey (2MASS), which is a joint project of the
University of Massachusetts and the Infrared Processing and Analysis
Center, funded by the National Aeronautics and Space Administration
and the National Science Foundation.   This publication makes use of
data products from the Wide-field Infrared Survey Explorer, which is a
joint project of the University of California, Los Angeles, and the
Jet Propulsion Laboratory/California Institute of Technology, funded
by the National Aeronautics and Space Administration. The WISE and
2MASS data are served by the NASA/IPAC Infrared Science Archive, which
is operated by the Jet Propulsion Laboratory, California Institute of
Technology, under contract with the National Aeronautics and Space
Administration.  This research has made use of the Digitized Sky
Surveys, which were produced at the Space Telescope Science Institute
under U.S. Government grant NAG W-2166.


\appendix

\section{Comparison to WISE}

\begin{figure}[H]
    \plotone{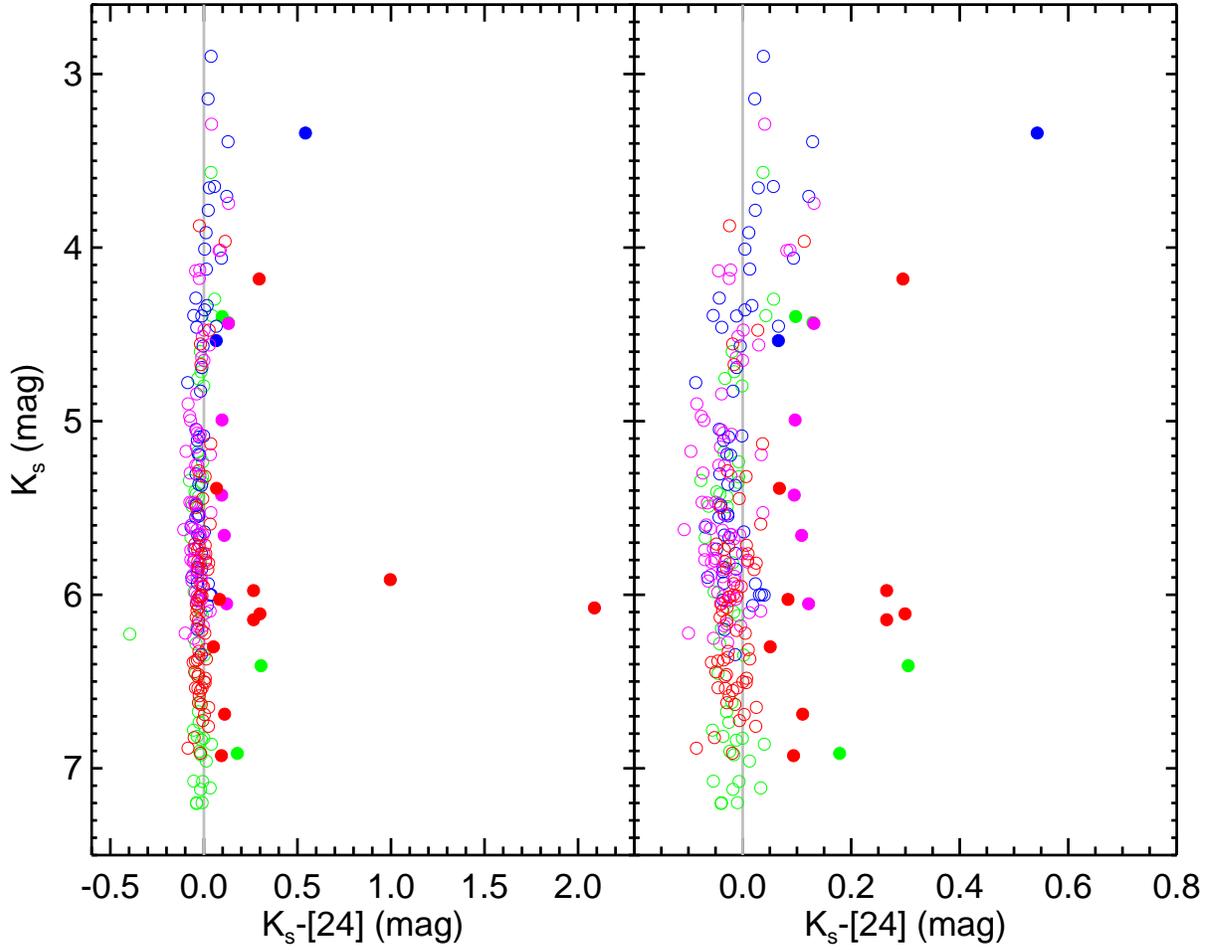}
\caption{$K_s$ vs. $K_s-[24]$ for our sample. The light grey line is
at $K_s-[24]$ = 0.  The left panel is full-scale; the right panel
zooms in on the $K_s-[24]$ = 0 regime. Open circles mark the
non-excess sample and closed circles mark the selected infrared
excesses as described in the text.  Green circles represent the blue
slow group, blue circles represent the blue rapid groups, magenta
circles represent the red slow group, and red circles represent the
red rapid group.  Note that the bulk of the distribution is not
centered on 0; see the text. The object with a large negative
excursion is also discussed in the text.  }
   \label{fig:kk24}
   \label{KvK24_full}
\end{figure}

\begin{figure}[H]
    \plotone{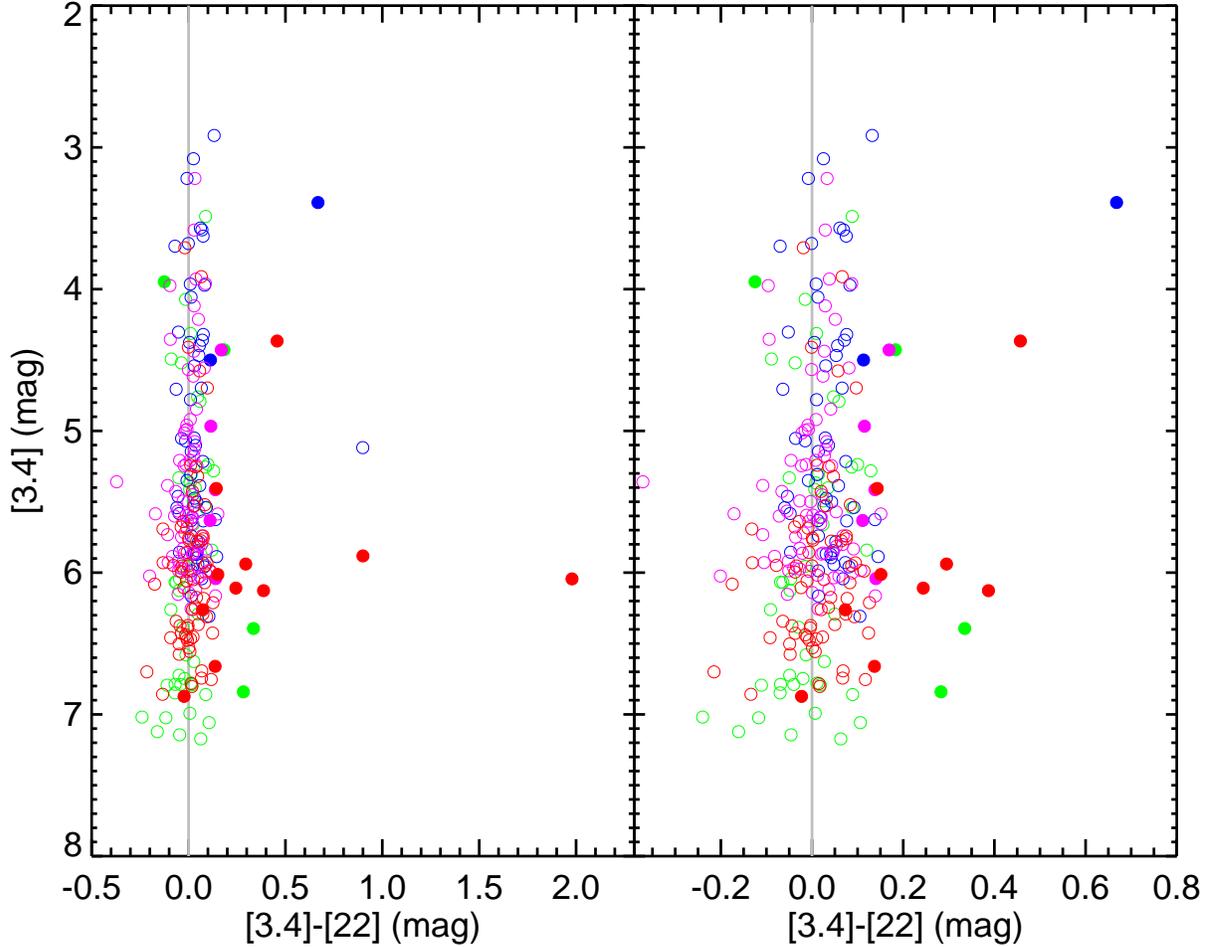}
\caption{$[3.4]$ vs. $[3.4]-[22]$ for our sample. The light grey line
is at $[3.4]-[22]$ = 0.  The left panel is full-scale; the right panel
zooms in on the $[3.4]-[22]$ = 0 regime.   Notation is as in previous
figure, where Open circles mark the non-excess sample and closed
circles mark the selected infrared excesses as described in the text.
The object selected as an IR excess with a large negative excursion
here (HIP 99680) and the object with [3.4]$-$[22]$>$0.5 that is not
selected as having an IR excess (HIP 3505) are identified as having
bad WISE measurements, and so are selected based on their $K_s-[24]$
values; see text. }
   \label{fig:w1w4}
\end{figure}

\begin{figure}[H]
    \plotone{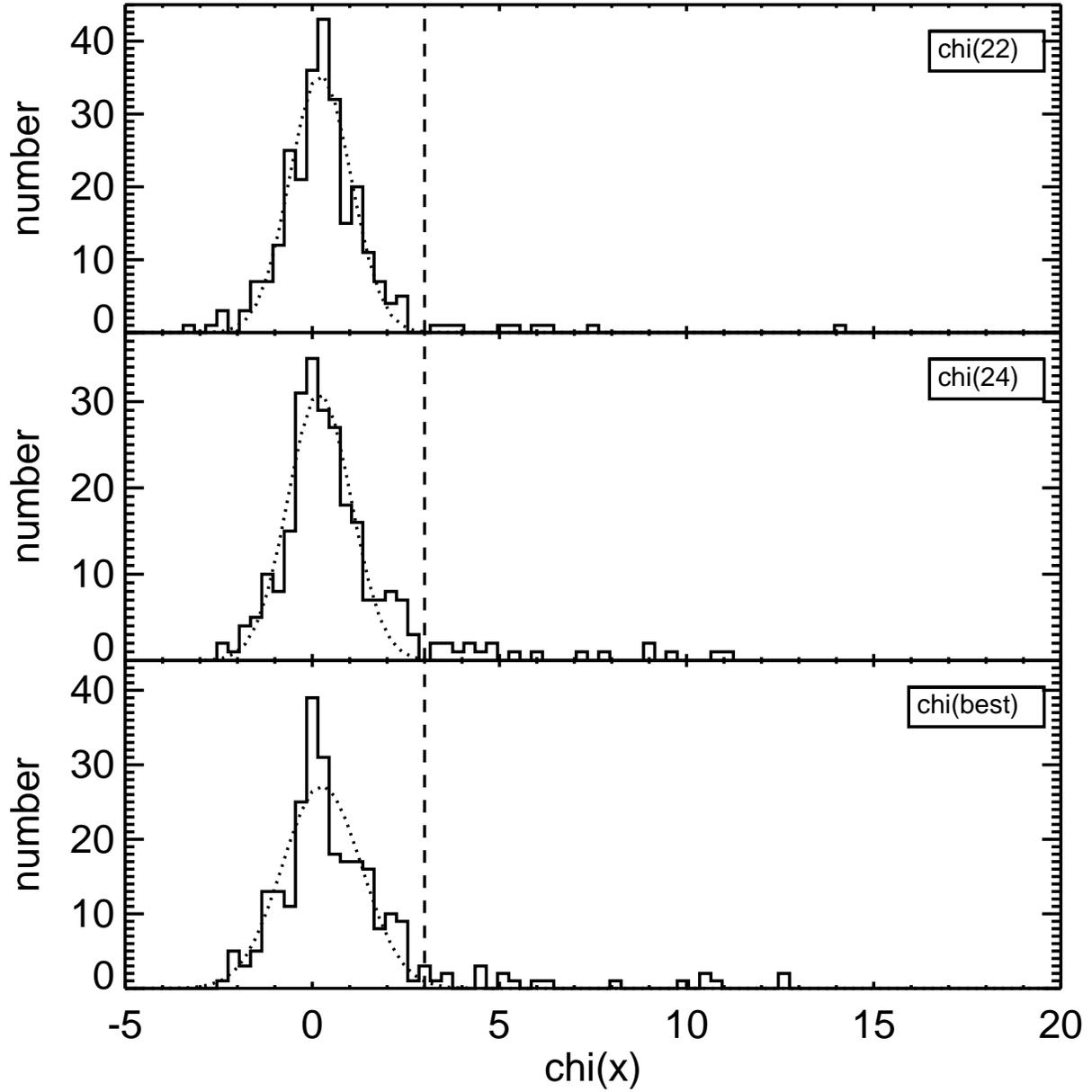}
\caption{Histograms of $\chi_{22}$ (top), $\chi_{24}$ (middle), and
$\chi_{best}$ (bottom) for
   the ensemble. Dotted lines are Gaussian fits to the distributions.
   Note that extreme negative and extreme positive values of $\chi$
   are not shown here.}
   \label{fig:chihisto}
\end{figure}

\begin{figure}[H]
    \plotone{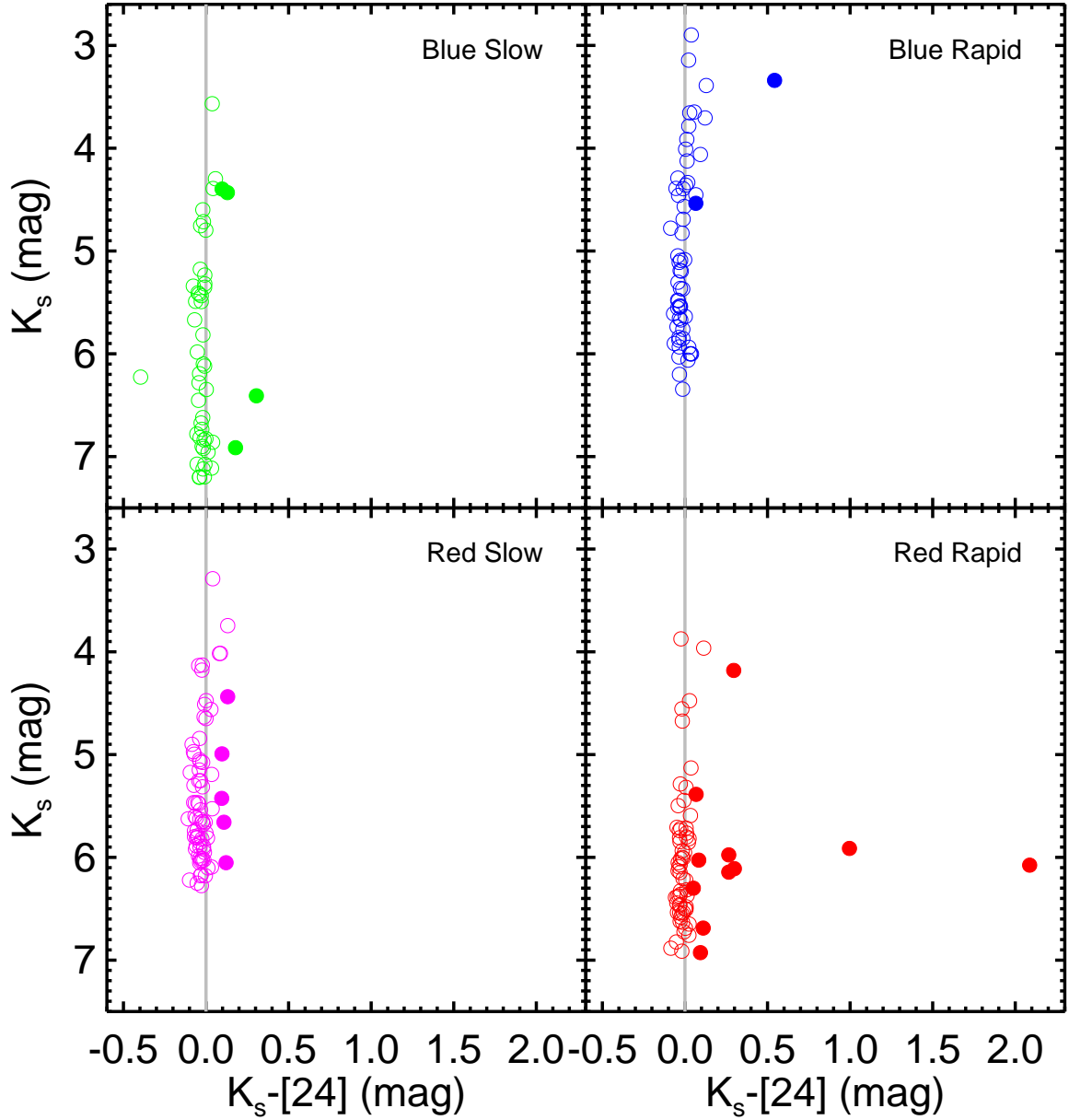}
\caption{$K_s$ vs. $K_s-[24]$ plot of each group individually and
their 24 \mum\ excesses; colors and symbols are as in previous figure.  The
most excess stars are found in the red rapid sample, but given small
number statistics, this is not significantly different than the other
samples.  See \S\ref{sec:disks} for discussion.}
   \label{KvK24_full_indiv}
\end{figure}

\begin{figure}[H]
    \plotone{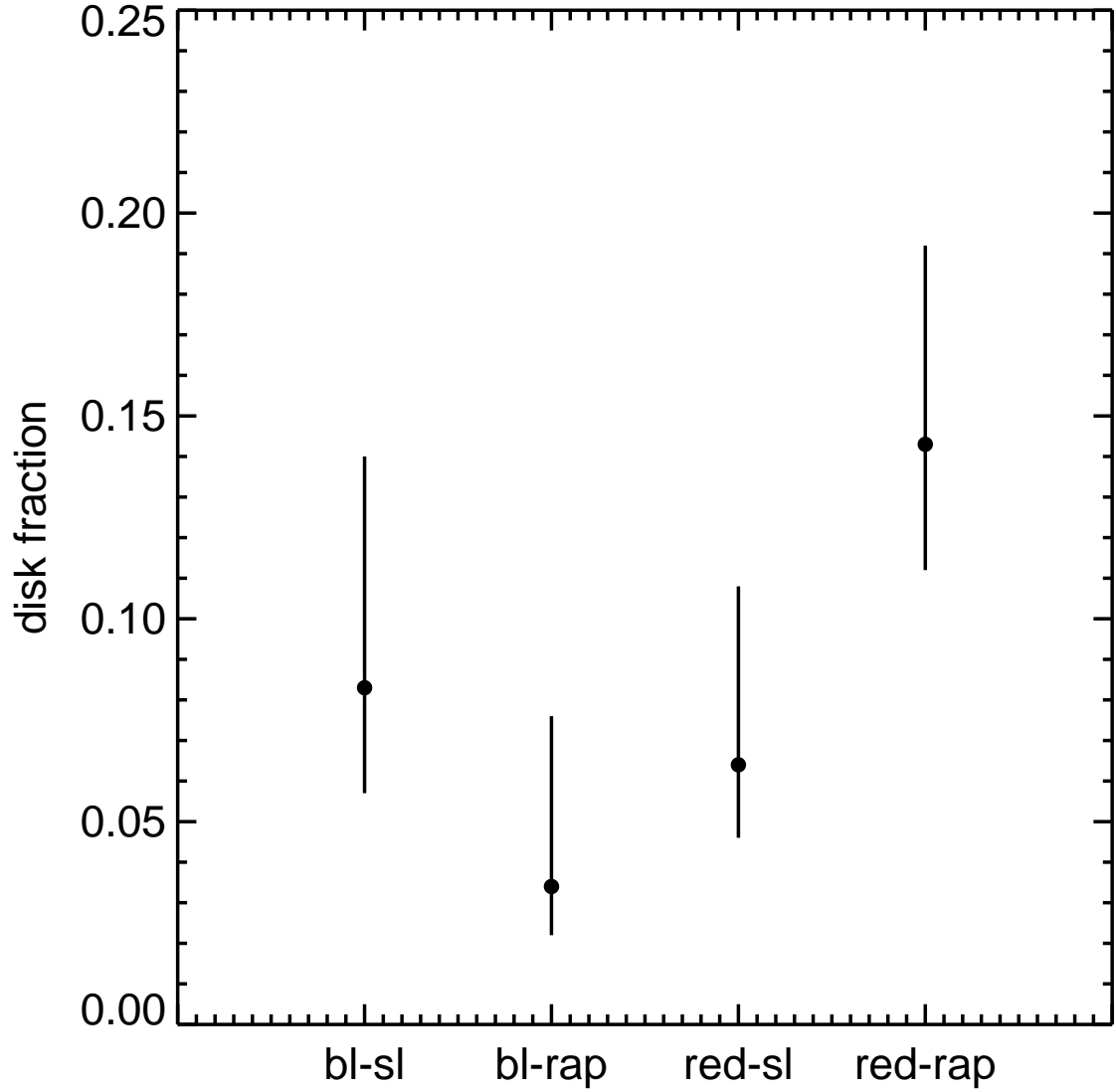}
\caption{Plot of disk fractions as calculated in this paper, with
formal errors calculated as per Burgasser \etal\ (2003).  The $x$-axis
is, from left to right, blue-slow, blue-rapid, red-slow, and
red-rapid.  The red-rapid sample has the largest fraction of disks,
and the blue-rapid sample has the smallest fraction of disks. }
   \label{fig:df}
\end{figure}

\begin{figure}[H]
    \plotone{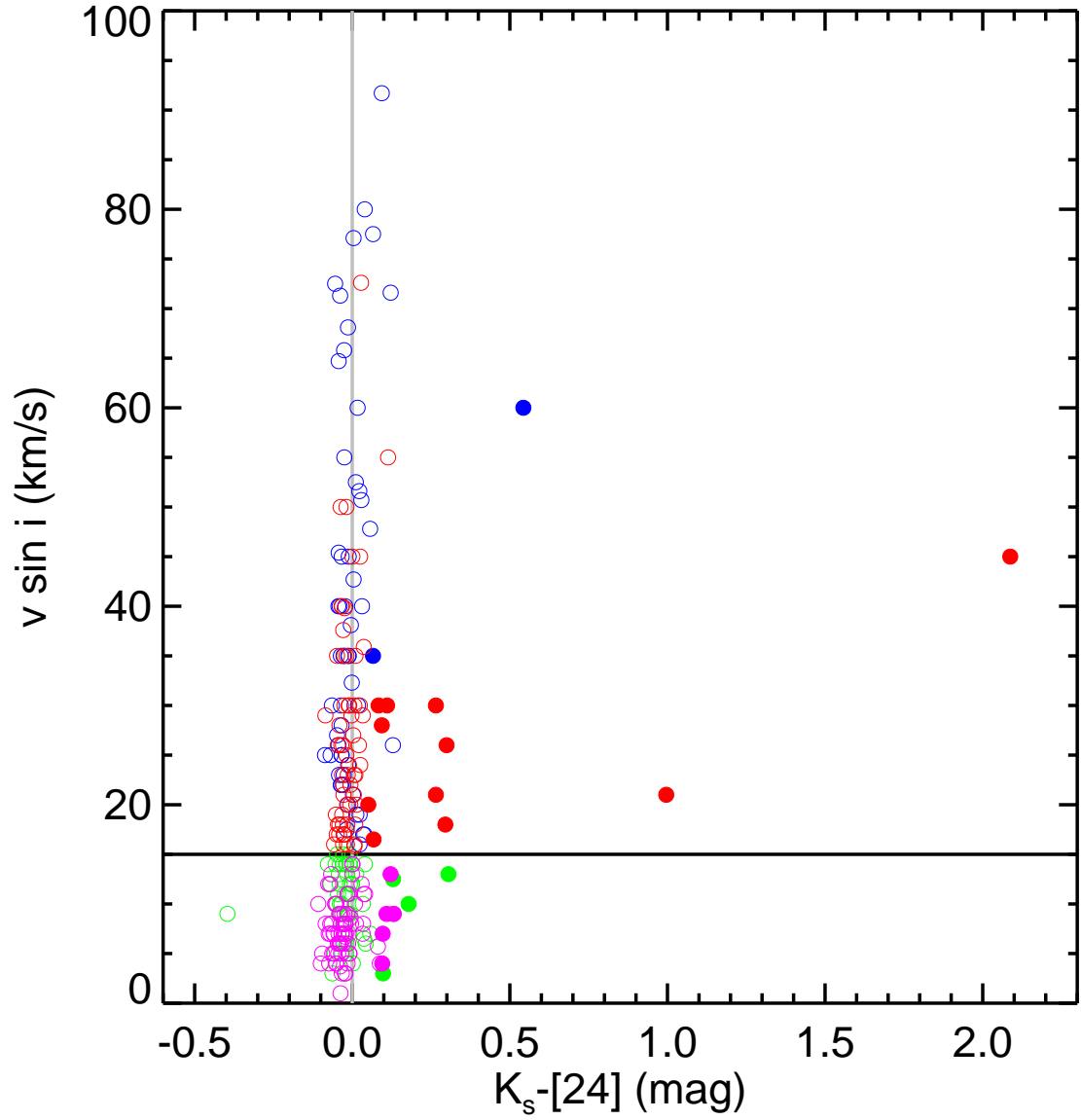}
\caption{\vsini vs.\ $K_s-[24]$ plot for our sample; notation is as in
previous figures.  This figure shows the range of rotation rates; the
division between slow and rapid rotators (indicated) is \vsini $\sim$
15 km s$^{-1}$.    }
  \label{vsini}
\end{figure}

\begin{figure}[H]
  \plotone{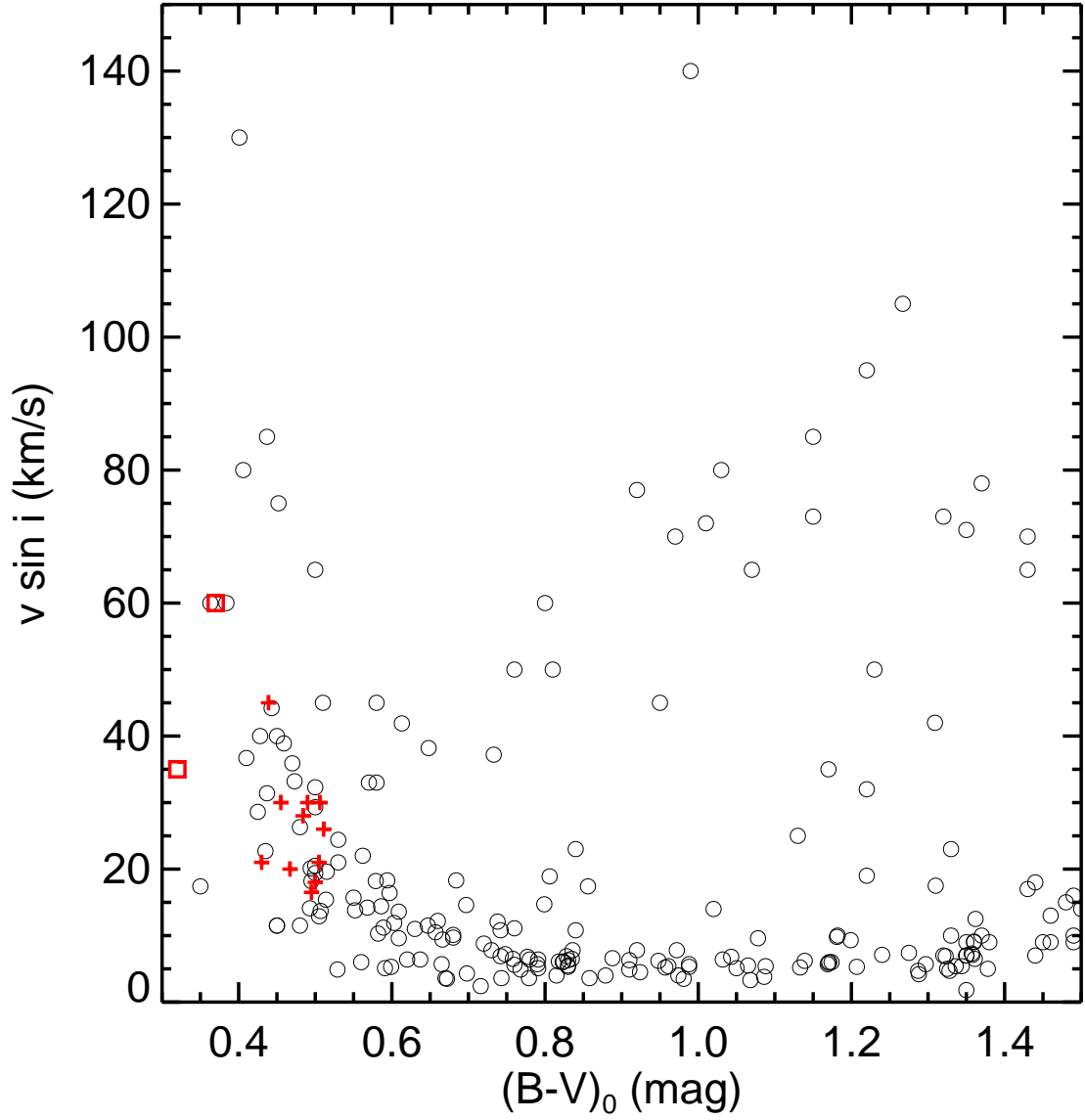}
\caption{$(B-V)_0$ vs.\ \vsini plot for the Pleiades (open circles), 
the eight red rapid rotators with disks from our analysis ($+$ signs),
and, for completeness, the two blue rapid rotators with an excess from
our analysis (squares). The rapid rotators from our analysis have
\vsini values consistent with those from `normal' F stars in the
Pleiades. }
  \label{vsiniandpleiades}
\end{figure}

This section explicitly considers a comparison of our photometric
results at 24 \mum\ to those from WISE at 22 \mum\ (from the all-sky
release from March 2012). As seen in Fig.~\ref{fig:kk24}, the ensemble
of disk-free stars is distributed about $K_s-[24]\sim0$, with a
slightly blue offset. The scatter around $K_s-[24]=-0.03$ (once
outliers are removed) is not very large. As seen in
Fig.~\ref{fig:w1w4}, the ensemble of disk-free stars is distributed
about $[3.4]-[22]=0$. Within WISE data, there is no offset, unlike
what we found in $K_s-[24]$. It is important to note that the 3.4
\mum\ (W1) detections here for the entire sample are saturated in the
original data, and the values here are obtained via PSF fitting to the
wings of the detection. The 22 \mum\ (W4) points are, in contrast, not
saturated. (The WISE Explanatory Supplement states that the channels
saturate at 8.1, 6.7, 3.8 and -0.4 mag for W1, W2, W3, and W4,
respectively, but that fits to the unsaturated wings of the PSF allow
viable magnitudes to be obtained up to 2.0, 1.5, -3.0 and -4 mag for
the four channels, respectively.)  A similar plot for this sample for
W2 is not at all as clean, as per the Explanatory Supplement (because
W2 is also saturated), but a similar plot for W3 is also very
well-behaved. Thus, the WISE data is quite internally consistent.

\begin{figure}[H]
    \plotone{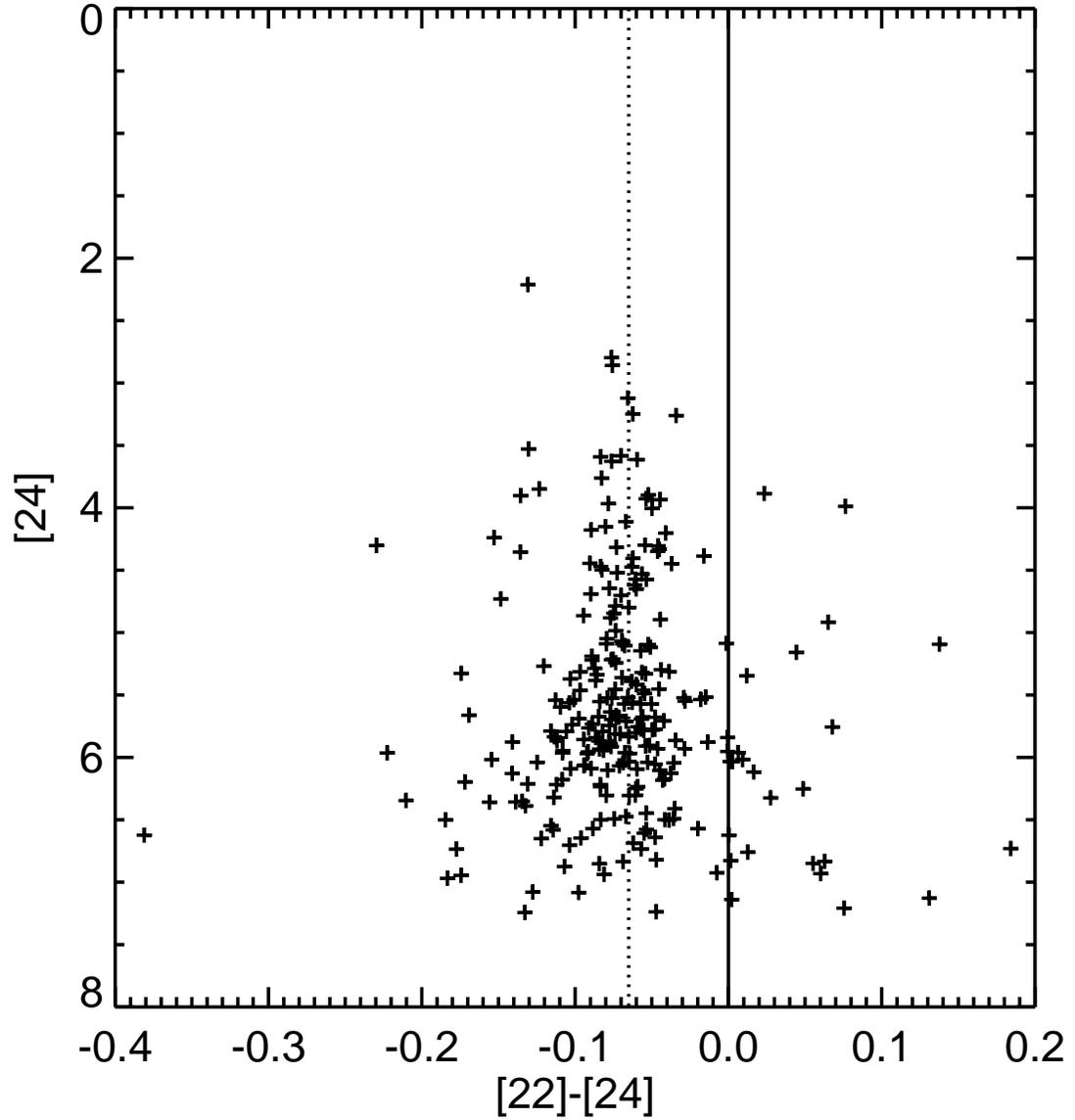}
\caption{$[24]$ vs.\ $[22]-[24]$ for our sample. Note that the mean of
the ensemble is $-$0.065 mags. There is evidently an offset between
the MIPS-24 zeropoint and the WISE-4 zero points for this sample.  }
   \label{fig:2224}
\end{figure}

Figure~\ref{fig:2224} explicitly compares the WISE-4 (22 \mum) points
and the MIPS-24 (24 \mum) points. In the ideal case, for plain
photospheres, the measured Vega-based magnitude of a star should be
the same at 22 and 24 \mum. Since we have all F stars here, both 22
and 24 \mum\ are on the Rayleigh-Jeans side of the SED, and as such,
the difference in magnitudes should be zero. As can be seen in
Figure~\ref{fig:2224}, it is not zero, but instead centered on
$-$0.065 mag. We wondered if this offset, while systematic, could be
well within expected errors; however, $>$95\% of the ensemble of [22]
errors for our sample as reported in the WISE catalog is less than
0.065 mags, and most of our MIPS-24 errors are $\sim$0.02 mags.  Our
targets are from all over the sky, with a wide variety of backgrounds,
but the backgrounds are still, on the whole, relatively low. The
source density is not very high; the sources are largely unconfused in
the MIPS data, and any source (e.g., apparent or actual binary, or
background galaxy) close enough in projected distance to adversely
affect the MIPS photometry should also affect the WISE photometry, in
the same way. The MIPS photometry, as well as the WISE photometry, is
obtained via PSF-fitting. The WISE photometry is calibrated primarily
from disk-free A stars; while color corrections are required for the
WISE photometry for very red sources, our targets should not require
such corrections.  The WISE team uses a different zero point for
MIPS-24; they use 7.449 Jy rather than the 7.17 Jy we used. The value
we use comes from the MIPS Instrument Handbook.  If we instead use
7.449 Jy, this {\em worsens} the discrepancy seen in
Figure~\ref{fig:2224}. 

\begin{figure}[H]
    \plotone{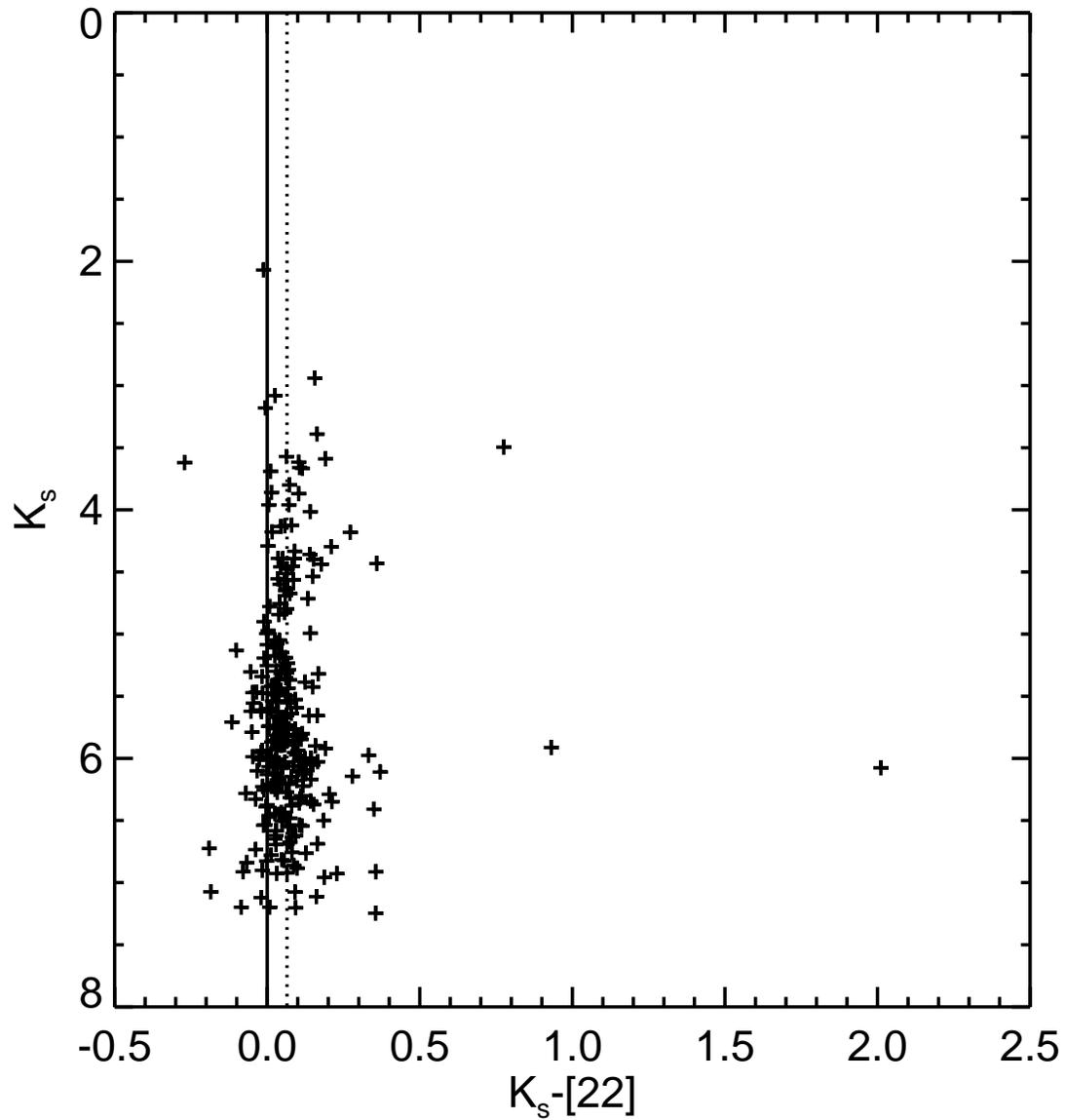}
\caption{$K_s$ vs.\ $K_s-[22]$ for our sample. Note that the mean of
the ensemble is not zero, but instead +0.065 (dotted line). }
   \label{fig:kk22}
\end{figure}

Figure~\ref{fig:kk22} shows $K_s$ vs.\ $K_s-[22]$ for our sample. As
in Fig.~\ref{fig:2224} above, there is an offset of 0.065 mag. Section
VI.3.i.4 of the Explanatory Supplement\footnote{Specifically,
http://wise2.ipac.caltech.edu/docs/release/allsky/expsup/sec6\_3c.html\#measure\_bias} 
includes a figure of [22] vs.\ $K_s-[22]$ for a cluster of stars;
there an offset can be seen in $K_s-[22]$ that is comparable to, or
even a little larger than, what we measure. So, there is a known
offset between $K_s$ and [22]. Since our [24] measurements are
well-matched to $K_s$ (as per Fig.~\ref{fig:kk24}), there is also an
offset between [22] and [24]; for our F star sample, this is
$\sim$0.065 mags. 

Absolute calibrations are difficult. While the WISE absolute
calibration may be correct in the end, the MIPS 24 \mum\ calibration
is closer to the 2MASS $K_s$ calibration, at least for our MIPS
reduction and our largely disk-free F star sample.

\clearpage

\begin{deluxetable}{lccccccccccccccc}
\tablecaption{Photometry and Ancillary Data\label{fulltab}}
\tabletypesize{\tiny}
\rotate
\tablewidth{0pt}
\tablehead{
\colhead{Name} & \colhead{Distance} &  \colhead{$V$} & 
\colhead{$B-V$} & \colhead{\vsini} & \colhead{sample} &
\colhead{$K_s$} &  \colhead{AORKey(s)} &\colhead{[24]} &
\colhead{$K_s-[24]$} &  \colhead{$\chi_{24}$} & 
\colhead{[3.4] (W1)} & \colhead{[22] (W4)} &
\colhead{[3.4]$-$[22]} &  \colhead{$\chi_{22}$}&  
\colhead{$\chi_{\rm best}$} \\
 & (pc) & mag & mag & km s$^{-1}$ &  & mag & & mag & mag & & mag & mag &
 mag  }
\startdata
   HIP462& 50& 7.13& 0.45& 24& Red-Rapid& 6.01$\pm$ 0.02&                      17346304&    6.017$\pm$ 0.022&   -0.010$\pm$ 0.030&   0.65& 5.95$\pm$ 0.05& 5.86$\pm$ 0.04&  0.087$\pm$ 0.066&              1.32&              1.14\\
   HIP606& 48& 7.05& 0.46&  9&  Red-Slow& 5.91$\pm$ 0.02&                      26198784&    5.930$\pm$ 0.022&   -0.015$\pm$ 0.029&   0.52& 5.86$\pm$ 0.05& 5.88$\pm$ 0.04& -0.021$\pm$ 0.066&             -0.32&              0.35\\
   HIP950& 21& 5.25& 0.44&  5&  Red-Slow& 4.13$\pm$ 0.31&                      10234112&    4.178$\pm$ 0.022&   -0.044$\pm$ 0.307&  -0.05& 4.12$\pm$ 0.09& 4.09$\pm$ 0.02&  0.029$\pm$ 0.095&              0.31&              0.28\\
  HIP1134& 46& 7.32& 0.50& 37& Red-Rapid& 6.07$\pm$ 0.02&                       5272064&    6.102$\pm$ 0.022&   -0.029$\pm$ 0.030&   0.04& 6.05$\pm$ 0.05& 6.02$\pm$ 0.05&  0.026$\pm$ 0.066&              0.40&              0.20\\
  HIP1402& 48& 6.45& 0.37&  8& Blue-Slow& 5.43$\pm$ 0.02&                      26213888&    5.464$\pm$ 0.022&   -0.029$\pm$ 0.028&   0.04& 5.40$\pm$ 0.06& 5.37$\pm$ 0.03&  0.035$\pm$ 0.072&              0.48&              0.21\\
  HIP1427& 46& 7.20& 0.48& 13&  Red-Slow& 6.05$\pm$ 0.02&    17951744,17951232,17951488&    5.931$\pm$ 0.022&    0.121$\pm$ 0.028&   5.49& 6.04$\pm$ 0.05& 5.90$\pm$ 0.05&  0.140$\pm$ 0.067&              2.08&              5.87\\
  HIP2337& 42& 6.14& 0.37& 40&Blue-Rapid& 5.20$\pm$ 0.02&                      26205184&    5.218$\pm$ 0.022&   -0.022$\pm$ 0.030&   0.26& 5.22$\pm$ 0.07& 5.14$\pm$ 0.03&  0.074$\pm$ 0.079&              0.94&              0.58\\
  HIP3214& 57& 7.59& 0.44& 30& Red-Rapid& 6.51$\pm$ 0.02&                      26180608&    6.499$\pm$ 0.023&    0.007$\pm$ 0.029&   1.26& 6.47$\pm$ 0.04& 6.46$\pm$ 0.05&  0.010$\pm$ 0.064&              0.16&              1.22\\
\enddata
\end{deluxetable}

\begin{deluxetable}{lcccc}
\tablecaption{Bright Sources}
\label{tab:saturatedphotometry}
\tablewidth{0pt}
\tablehead{
\colhead{Name} & \colhead{$K_{s}$ from 2MASS} & 
\colhead{$K_{s}$ used here} & \colhead{$K_s-[24]$} &
\colhead{[3.4]$-$[22] } }
\startdata
HIP16245 & 3.936 & 3.79 &   0.02 &   0.00\\ 
HIP21770 & 3.724 & 3.64 &   0.06 &   0.06\\ 
HIP22449 & 1.600 & 2.11 &  -0.10 &   0.22\\ 
HIP28103 & 2.993 & 2.90 &   0.04 &   0.13\\ 
HIP33202 & 3.888 & 3.91 &   0.01 &  -0.07\\ 
HIP34834 & 3.320 & 3.67 &   0.03 &   0.07\\ 
HIP36366 & 2.978 & 3.39 &   0.13 &  -0.01\\ 
HIP38423 & 3.841 & 3.96 &   0.11 &  -0.02\\ 
HIP40843 & 3.868 & 4.02 &   0.08 &   0.04\\ 
HIP50954 & 3.252 & 3.14 &   0.02 &   0.02\\ 
HIP61174 & 3.372 & 3.34 &   0.54 &   0.67\\ 
HIP63503 & 3.953 & 4.06 &   0.09 &   0.08\\ 
HIP64241 & 3.108 & 3.29 &   0.04 &   0.03\\ 
HIP73996 & 3.863 & 3.87 &  -0.02 &   0.07\\ 
HIP76829 & 3.802 & 3.71 &   0.12 &   0.07\\ 
HIP80179 & 4.094 & 4.01 &   0.00 &   0.01\\ 
HIP82860 & 3.736 & 3.75 &   0.13 &   0.03\\ 
HIP96441 & 3.537 & 3.57 &   0.04 &   0.09\\ 
\enddata
\label{brightk}
\end{deluxetable}

\begin{deluxetable}{lcl}
\tablecaption{Stars with IR excesses or needing other special notes \label{tab:disks}}
\tablewidth{0pt}
\rotate
\tabletypesize{\tiny}
\tablehead{
\colhead{Name} & \colhead{Has IR excess here?} &  \colhead{notes} }
\startdata
HIP1427 & Yes&
$\chi_{24}>3,\chi_{22}<3$ but $\chi_{\rm best}>3$; new IR excess identified here
\\
HIP3505 & No&
one of the WISE bands is evidently wrong, so taking $\chi_{\rm best}=\chi_{24}$ 
\\
HIP4868 & No&binary\\
HIP5141 & No&binary\\
HIP5144 & No&binary\\
HIP5631 & Yes&
$\chi_{24}>3,\chi_{22}<3$ but $\chi_{\rm best}>3$; new IR excess identified here
\\
HIP7916 & No&binary\\
HIP8514 & No&binary\\
\enddata
\end{deluxetable}

\begin{deluxetable}{lcccc}
\tablecaption{Excess Results\label{excess}}
\tablewidth{0pt}
\tablehead{
\colhead{Group Name} & \colhead{\# of stars} &  \colhead{\# of excesses} 
& \colhead{excess fraction} & \colhead{1$\sigma$ errors} }
\startdata
Blue Slow  & 49 &  4  & 0.083  & 0.057 to 0.140  \\
Blue Rapid & 59 &  2  & 0.034  & 0.022 to 0.076  \\
Red Slow   & 78 &  5  & 0.064  & 0.046 to 0.108  \\
Red Rapid  & 77 & 11  & 0.143  & 0.112 to 0.192  \\
\enddata
\end{deluxetable}

\end{document}